\documentclass{aastex62}

\usepackage{hyperref}
\submitjournal{ApJ Supplement}

\shorttitle{Self-Gravitational Force Calculation for AMR}
\shortauthors{Tseng et al. (2019)}
\begin{document}

\title{Direct Calculation of Self-Gravitational Force for Infinitesimally Thin Gaseous Disks Using Adaptive Mesh Refinement}

\correspondingauthor{Chien-Chang Yen}
\email{yen@math.fju.edu.tw}

\author[0000-0002-0786-7307]{Yao-Huan Tseng } %（曾耀寰）}
\affiliation{Institute of Astronomy and Astrophysics, Academia Sinica, Taipei 10617, Taiwan, R.O.C.}
\affiliation{Academia Sinica, Theoretical Institute for Advanced Research in Astrophysics}

\author[0000-0001-8385-9838]{Hsien Shang }% （尚賢）}
\affiliation{Institute of Astronomy and Astrophysics, Academia Sinica, Taipei 10617, Taiwan, R.O.C.}
\affiliation{Academia Sinica, Theoretical Institute for Advanced Research in Astrophysics}

\author{Chien-Chang Yen } %（嚴健彰）}
\affiliation{Department of Mathematics, Fu Jen Catholic University, New Taipei City, Taiwan}
\affiliation{Institute of Astronomy and Astrophysics, Academia Sinica, Taipei 10617, Taiwan, R.O.C.}
\affiliation{Academia Sinica, Theoretical Institute for Advanced Research in Astrophysics}

\begin{abstract}

Yen et al. (2012) advanced a direct approach for the calculation of self-gravitational force to second order accuracy based on uniform grid discretization. This method improves the accuracy of $N$-body calculation by using exact integration of kernel functions and employing the Fast Fourier Transform (FFT) to reduce complexity of computation to nearly linear. This direct approach is free of artificial boundary conditions, however, the applicability is limited by the uniform discretization of grids. We report here an advancement in the direct method with the implementation of adaptive mesh refinement (AMR) and maintaining second-order accuracy, which breaks the barrier set by uniform grid discretization. The adoption of graphic process units (GPUs) can significantly speed up the computation and make application of this method possible to astrophysical systems of gaseous disk galaxies and protoplanetary disks. 
\end{abstract}


\section{Introduction} 
\label{sec:intro}

Self-gravity of gas is significant in many areas of astrophysics. Dynamical evolution of the center region of galaxies, such as the starburst rings (Lin et al. 2013; Seo \& Kim 2014), the sub-parsec scale within the Active Galactic Nuclei (AGN), inner bars and spirals (Kim et al. 2012; Lee \& Shu 2012; Lee 2014), and bright and young stars formed along spiral arms (Elmegreen et al. 2014), are consequences of self-gravitating interactions of gas. Such phenomena exist because the interstellar gas has to be transferred to the central region at a typical scale of 10-kpc. Meanwhile, the self-gravity of gas play an important role to the evolution of gas in disk galaxies. A disk is regarded to be thin relative to the physical size scale when
the ratio of lengths of vertical and radial directions is around 0.001-- 0.01. An infinitesimally thin gaseous disk represents the disk as a distribution of two dimensional surface density in three dimensional space.

The potential $\Phi$ for a given distribution of gaseous density $\rho$ in three-dimensional space satisfies the Poisson equation: 
\begin{eqnarray}
\nabla\cdot (\nabla \Phi)(x,y,z)=4\pi G\rho (x,y,z),
\label{eqnPhi}
\end{eqnarray}
where $\nabla =(\frac{\partial}{\partial x},\frac{\partial}{\partial y},\frac{\partial}{\partial z})$,
or as an integral 
\begin{eqnarray*}
\Phi(x,y,z)=-G\int\!\!\!\int\!\!\!\int 
K(\bar x -x, \bar y-y, \bar z-z)\rho (\bar x,\bar y,\bar z)\, d\bar x\,d\bar y \,d\bar z
\end{eqnarray*}
where $\displaystyle K(x,y,z)=\frac{1}{\sqrt{x^2+y^2+z^2}}$ and $G$ is the gravitational constant. 
The density $\rho(x,y,z)$ for a gaseous disk is associated with a
surface density $\sigma(x,y)$ by 
\begin{eqnarray}
\rho(x,y,z)=\sigma(x,y)\delta(z),
\label{eqnrho}
\end{eqnarray}
where $\delta$ is the Dirac function.
Furthermore, the potential in mid-plane, $\Phi(x,y,0)$, is associated with a singular kernel integral through
\begin{eqnarray*}
\Phi(x,y,0)=-G\int\!\!\!\int
K(\bar x-x,\bar y-y,0)\sigma(\bar x, \bar y)\,
d\bar x\,d\bar y.
\end{eqnarray*}

It is well known multigrid method can solve the potential, which satisfies Eq.~(\ref{eqnPhi}) on finite region with some prescribed boundary conditions with linear complexity. In three dimension, the linear complexity is  $O(N^3)$
where $N$ is the number of zones in one dimension. In the current situation, the boundary is open and the values at boundary are required to be evaluated. The  complexity will reach $O(N^5)$ using direct approach and it can be reduced to $O(N^4)$ if the density is replaced by surface density $\sigma(\bar x, \bar y)$
defined in Eq.~(\ref{eqnrho}).
Another evaluation of potential to deal with vaccum boundary is developed by James (1977) and it is extended to cylindrical coordinates (Moon et al. 2019). Once the potential obtained, one can take the numerical derivatives to get the forces. Therefore, the second order of accuracy for force calculation should be guaranteed by more than second order of accuracy for potential (Gupta et al. (1997); Lai \& Tseng (2007)).

A slef-gravity calculation can use the differential equation or the integral form. An integral form is more efficient than a differential equation for an infinitesimally thin gaseous disk. A direct calculation of the force in the $x$-direction is to compute the derivative
\begin{eqnarray}
\label{eqnXforce}
-\frac{\partial}{\partial x}\Phi(x,y,0)
=-\int\!\!\!\int \frac{\partial}{\partial x}K(\bar x-x,y-y,0)\sigma(\bar x,\bar y)\,
d\bar x\,d\bar y,
\end{eqnarray}
and similarly, the calculation of $y$-force and $z$-force are to compute the $y$- and $z$-partial
derivative of the potential, respectively.
A direct method for self-gravitational force calculation as shown in Eqn (\ref{eqnXforce}) has been proposed in Yen et al. (2012).
 This approach can be regarded as an improvement to the $N$-body calculation on order of accuracy using exact integration of the kernel functions. A fast computation for the method based on uniform grid discretization was developed by employing the fast Fourier transform (FFT), which reduces the complexity of computation to nearly linear. This approach has also been shown to second order accuracy. It was fast, and free of artificial boundary conditions.
Such properties of the method have been developed and demonstrated for the uniform grid.
Moreover, it has been extended for the nested-grids in Cartesian coordinates by Wang et al. (2016). 
Unfortunately, the order of accuracy for this approach on polar coordinate using logarithmic discretization in radial direction will be reduced due to the fact that no analytic formula exists for the integral of logarithmic function. For preserving second order of accuracy, current work is limited to the Cartesian coordinates.

Adaptive mesh refinement (AMR) in hydrodynamic simulations has become a highly desirable target for the generalization of self-gravitational force calculations. Here, we consider computation of self-gravity with AMR by preserving the order of accuracy. The technique of FFT itself in this case, however, is unable to accelerate the computation for the increased complexity. To carry out the full fast calculation with a direct method, one would need to rely on recent improvement of computational power.
More precisely, we employ the graphic process units (GPU) in this new development to reduce the computational time. In short, we generalize Wang et al. (2016) to accommodate the grid generation in AMR with the computational capability of GPUs at state-of-art level. Meanwhile, we develop two processes of level refinement: independent and variable, for which the computations will be carried out and benchmarked.

This paper is organized as follows. We first outline generalization of Yen et al. (2012) into the AMR grid in Section 2. We carry out simulations using both uniform and AMR grid discretization with independent and variable level refinement approaches to investigate accuracy, order of accuracy, and the computational time in reality in Section 3.  We conclude this work in the last section. 

\section{Direct Method Based on AMR} \label{sec:A direct method}

The proposed direct method of force calculation based on AMR is more flexible for grid discretization, conceptually simpler, and straightforward to implement. Let us start with a surface density distribution $\sigma(x,y)$, which is assumed to vanish outside of a rectangular region $D$, and
\begin{eqnarray}
\label{eqnSigma}
\sigma =\sum^{\cal N}_{i=1}\sigma_i(x,y), 
\end{eqnarray}
where $\sigma_i(x,y)=\sigma(x,y)\chi_{D_i}$, $\chi_{D_{i}}$ is a characteristic function of the domain $D_i$,
\begin{eqnarray*}
\chi_{D_i}(x,y)
=\left\{
\begin{array}{ll}
1, &\quad (x,y)\in D_i,\\
0, &\quad (x,y)\not \in D_i,
\end{array}
\right.
\end{eqnarray*}
with the sub-rectangle $D_{i}$ mutually disjoint with each other except at the boundary and $\displaystyle \bigcup^{\cal N}_{i=1} D_{i}=D$. The positive integer ${\cal N}$ is the total number of zones generated by the AMR grids. From Eqs. (\ref{eqnXforce}) and (\ref{eqnSigma}), it follows that 
\begin{eqnarray}
\label{eqnXforce2}
-\frac{\partial}{\partial x}\Phi(x,y,z)
=\sum^{\cal N}_{i=1}  \int\!\!\!\int_{D_i} 
\frac{\partial}{\partial x} (K(\bar x-x,\bar y-y,z) )
\sigma_{i}(\bar x,\bar y)\,d\bar x\,d\bar y.
\end{eqnarray}
In Yen et al (2012), the restricted surface function 
$\sigma_{i}$ on $D_i$ is approximated by 
a Taylor polynomial of order 1, and it is given as 
\begin{eqnarray}
\label{eqnGamma}
\sigma_{i}(x,y) \approx
\sigma^0_{i} + \delta^x_{i}(x-x^c_i)
+\delta^y_{i}(y-y^c_i) 
\end{eqnarray}
in the subrectangle region $D_{i}$, with
$(x^c_i,y^c_i)$ being the center of the sub-rectangle $D_{i}$. Substituting equations (\ref{eqnGamma}) into (\ref{eqnXforce2}) gives us
\begin{eqnarray*}
-\frac{\partial}{\partial x}\Phi(x,y,z)
&\approx&-\sum^{\cal N}_{i=1}  \int\!\!\!\int_{D_i} 
\frac{\partial}{\partial x} (K(\bar x-x,\bar y-y,z) )
\left[
\sigma^0_{i} + \delta^x_{i}(x-x^c_i)
+\delta^y_{i}(y-y^c_i) 
\right]
\,d\bar x\,d\bar y\\
&=& 
-\sum^{\cal N}_{i=1}\sigma^0_{i}
 \int\!\!\!\int_{D_i} 
\frac{\partial}{\partial x} (K(\bar x-x,\bar y-y,z) )
\,d\bar x\,d\bar y\\
&-&
\sum^{\cal N}_{i=1} \delta^x_{i}\int\!\!\!\int_{D_i} 
\frac{\partial}{\partial x} (K(\bar x-x,\bar y-y,z) )
(\bar x-x^c_i)
\,d\bar x\,d\bar y\\
&-&
\sum^{\cal N}_{i=i} \delta^y_{i}\int\!\!\!\int_{D_i} 
\frac{\partial}{\partial x} (K(\bar x-x,\bar y-y,z) )
(\bar y-y^c_i) 
\,d\bar x\,d\bar y.\\
\end{eqnarray*}

We define $F^x_i(x,y,z)=-\frac{\partial}{\partial x}\Phi(x,y,z)$
and we obtain
\begin{eqnarray*}
F^x_i (x,y,z)\approx F^{x,0}_i (x,y,z)+F^{x,x}_i (x,y,z)+F^{x,y}_i(x,y,z)
\end{eqnarray*}
where
\begin{eqnarray*}
F^{x,0}_i (x,y,z)& = & -\sum^{\cal N}_{i=1} \sigma^0_i K^{x,0}_i(x,y,z),\\
F^{x,x}_i (x,y,z)& = & -\sum^{\cal N}_{i=1} \delta^x_i K^{x,x}_i(x,y,z),\\
F^{x,y}_i (x,y,z)& = & -\sum^{\cal N}_{i=1} \delta^y_i K^{x,y}_i(x,y,z),\\
\end{eqnarray*}
and 
\begin{eqnarray*}
K^{x,0}_i(x,y,z) & = & \int\!\!\!\int_{D_i}\frac{(\bar x - x)}{((\bar x-x)^2+(\bar y-y)^2+z^2)^{3/2}}\,
d\bar x\,d\bar y, \\
K^{x,x}_i(x,y,z) & = & \int\!\!\!\int_{D_i}\frac{(\bar x - x^c_i)(\bar x-x)}{((\bar x-x)^2+(\bar y-y)^2+z^2)^{3/2}}\,
d\bar x\,d\bar y, \\
K^{x,y}_i(x,y,z) & = & \int\!\!\!\int_{D_i}\frac{(\bar y - y^c_i)(\bar x-x)}{((\bar x-x)^2+(\bar y-y)^2+z^2)^{3/2}}\,
d\bar x\,d\bar y. \\
\end{eqnarray*}

We further decompose the sub-rectangle domain $D_i=[x^\ell_i,x^r_i]\times [y^d_i,y^u_i]$. 
The full expression of force kernels that are summarized as below and was formulated in Yen et al. (2012) when $z=0$, using $[x^\ell_i,x^r_i]$ and $[y^d_i,y^u_i]$:
\begin{eqnarray*}
K^{x,0}_i(x,y,z)&=&-\ln ((\bar y-y)+\sqrt{(\bar x-x)^2+(\bar y-y)^2+z^2})|^{x^r_i}_{x^\ell_i}|^{y^u_i}_{y^d_i},\\
K^{x,x}_i(x,y,z)&=&(x-x^c_i) K^{x,0}_i(x,y,z) 
+\left[(\bar y-y)\ln ((\bar x-x)+\sqrt{(\bar x-x)^2+(\bar y-y)^2+z^2})\right.\\
&&\left.-|z|sgn(x)\arctan (\frac{sgn(x)xy}{|z|\sqrt{(\bar x-x)^2+(\bar y-y)^2+z^2}})|^{x^r_i}_{x^\ell_i}|^{y^u_i}_{y^d_i}\right],\\
K^{x,y}_i(x,y,z)&=&( y-y^c_i)K^{x,0}_i(x,y,z)
-\left[\sqrt{(\bar x-x)^2+(\bar y-y)^2+z^2}|^{x^r_i}_{x^\ell_i}|^{y^u_i}_{y^d_i}\right],\\
K^{y,0}_i(x,y,z)&=&-\ln ((\bar x-x)+\sqrt{(\bar x-x)^2+(\bar y-y)^2+z^2})|^{x^r_i}_{x^\ell_i}|^{y^u_i}_{y^d_i},\\
K^{y,x}_i(x,y,z)&=&( x-x^c_i)K^{y,0}_i(x,y,z)
-\left[\sqrt{(\bar x-x)^2+(\bar y-y)^2+z^2}|^{x^r_i}_{x^\ell_i}|^{y^u_i}_{y^d_i}\right].\\
K^{y,y}_i(x,y,z)&=&(y-y^c_i) K^{y,0}_i(x,y,z)
+\left[(\bar x-x)\ln ((\bar y-y)+\sqrt{(\bar x-x)^2+(\bar y-y)^2+z^2})\right.\\
&&\left.-|z|sgn(y)\arctan
(\frac{sgn(y)xy}{|z|\sqrt{(\bar x-x)^2+(\bar
y-y)^2+z^2}})
|^{x^r_i}_{x^\ell_i}|^{y^u_i}_{y^d_i}\right],\\
K^{z,0}_i(x,y,z)&=&
\arctan(\frac{xy}{z\sqrt{(\bar x-x)^2+(\bar y-y)^2+z^2}})|^{x^r_i}_{x^\ell_i}|^{y^u_i}_{y^d_i}\\
K^{z,x}_i(x,y,z)&=&
(x-x^c_i)K^{z,0}_i(x,y,z)
+zK^{x,0}_i(x,y,z)\\
K^{z,y}_i(x,y,z)&=&
(y-y^c_i)K^{z,0}_i(x,y,z)
+zK^{y,0}_i(x,y,z)\\
\end{eqnarray*}
This full expression can be evaluated if the position $(x,y)$ does not coincide with the position $\bar x =x$ or $\bar y=y$ on the plane, $z=0$. For uniform grid discretization, the center of the zone, $\bar x$ and $\bar y$ will never fall on the interface; however, this situation may not be completely avoidable in the context of nested grid (Wang et al. 2016) or the AMR. When such situation occurs, the Kernel $K^{x,0}_i(x,y,0)$ can be modified to be 
\begin{eqnarray*}
-\mbox{sgn}(|\bar x-x|)\ln ((\bar y-y)+
\sqrt{(\bar x-x)^2+(\bar y-y)^2})
-(1-\mbox{sgn}(|\bar x-x|))\mbox{sgn}(\bar y-y)\ln |\bar y-y|,
\end{eqnarray*}
which can still be used for the uniform grids but evaluating it would take more computational time. 

\section{Numerical Simulations}

\label{sec:Numerical Simulations}

In this section, we carry out simulations involving self-gravitational force calculation in thin disks based on the AMR formulation derived above. The force is calculated at the center of each zone. We will demonstrate that our updated direct method for the calculation of self-gravitational force fits well under the framework of AMR, with two refinement approaches. One is the Independent Level Refinement (ILR) in which each level refined is independent of the value of the level, and the other is the Variable Level Refinement (VLR) in which each of the refinement taking place is variable with the value of the level.

For simplicity, the domain of consideration is confined to a rectangle for the coarsest grid discretized. Each refinement subdivides each parent zone into four sub-rectangles. We record each cell by a tuple of $9$ elements, ($x_c$, $y_c$, L, P, $EL$, LN, DN, RN, UN), whose center $(x_c,y_c)$ is located at level $L$ for each cell, with parent $P$ and the end of leaf $EL$. The elements LN, DN, RN, and UN, are the nearby nodes on the sides of left, down, right, and up. This generation of grid allows an easier implementation into the hydrodynamical simulations.

In the hydrodynamical simulation, one needs to consider the grid zones at the same and different levels for the calculation of flux. Here we concentrate on the self-gravitational force calculation, which is only restricted to the leaf nodes (zones). We apply our approach to an AMR grid generated in the sense of Berger \& Oliger (1984), as shown Figure 1. The generation of an uniform grid is initiated at the coarsest level $L=1$, where the coarsest grid is divided into (1, 2, 3, ...16), a total of 16 cell zones. The first refinement is taking place at the cell zones $6$ and $11$ to the level of $L=2$, where the cell zone $6$ is divided into zones (17, 18, ,19, 20) and zone $11$ is subdivided into zones (21, 22, 23, 24). For $L=3$ refinement within zones $17$ and $19$, they are referred to as zones (25, 26, 27, 28) and (29, 30, 31 32), respectively. We report the complete list of recorded data in Table 10 of the Appendix A. 

\begin{figure}[h]
\centerline{\includegraphics[scale=.3]{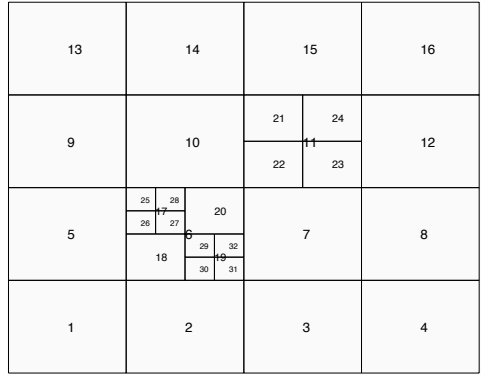} }
%\epsscale{0.6}
%\plotone{AMRG.eps}
\caption{Illustration of the adaptive mesh refinement to $L=3$ level starting from a coarsest level of 16 zones.}
\end{figure} 
%%%

We first address the preservation of second order accuracy in the complexity of computation on an AMR grid.  We adopt the $D_2$ disk in the family of finite disk with surface density $\Sigma_n (R; \alpha)= (1-R^2/\alpha^2)^{n-1/2}$ from Schulz (2009), for which $n=2$.  The $D_2$ disk has the following surface density:
\begin{eqnarray}
\Sigma_{D_2}(R;\alpha)=\left\{
\begin{array}{ll} \sigma_0 (1-R^2/\alpha^2)^{3/2}& \mbox{ for } R<\alpha,\\
0 & \mbox{ for } R>\alpha.
\end{array}
\right.
\end{eqnarray}

The corresponding potential in explicit form at the mid-plane $z=0$ as shown in Eqn. (29) of Schulz (2009) is, 
\begin{eqnarray}
\Phi_{D_2}(R,0;\alpha)=\left\{
\begin{array}{ll}-\frac{3\pi^2\sigma_0 G}{64\alpha^3}(8\alpha^4-8\alpha^2R^2+3R^4)&
\mbox{ for } R\le \alpha,\\
-\frac{3\pi\sigma_0 G}{32\alpha^3}\left[(8\alpha^4-8\alpha^2R^2+3R^4)\sin^{-1}
(\frac{\alpha}{R})+3\alpha(2\alpha^2-R^2)\sqrt{R^2-\alpha^2}\right]& \mbox{ for } R>\alpha,
\end{array}
\right.
\end{eqnarray}
and the radial force can be found by differentiating the potential with respect to $R$ (Eqn. [30] of Schulz 2009) as
\begin{eqnarray}
F_{R,D_2}(R,0;\alpha)=\left\{
\begin{array}{ll}-\frac{3\pi^2\sigma_0 RG}{16\alpha^3}(4\alpha^2-3R^2)&
\mbox{ for } R\le \alpha,\\
-\frac{3\pi\sigma_0 G}{8\alpha^3}\left[R (4\alpha^2-3R^2)\sin^{-1}(\frac{\alpha}{R})
-\alpha(2\alpha^2-3R^2)\sqrt{1-\alpha^2/R^2}\right]& \mbox{ for } R>\alpha.
\end{array}
\right.
\end{eqnarray}

With the help of the potential-density pair, we compare the numerical results with the analytic calculations. The $D_2$ disk is not smooth at its boundary, therefore it is suitable for a study based on the AMR approach to demonstrate the resulting improvement. This is a feature that can not be explicitly shown by smooth functions either with a uniform or adaptively refined grid setting. Since the $D_2$ disk is symmetric, another case is a $D_{2,2}$ disk which consists of two $D_2$ disks which have the centers located at $(0,0.5)$ and $(0,-0.5)$ of $D_2$, respectively, with $\alpha=0.25$ breaks the property of symmetry. It is regarded as a general situation.  

Without loss of generality, order of accuracy and computational complexity, we set the computational domain at 
$\Omega=[-1,1]\times [-1,1]$, 
$\sigma_0=G=1$ 
and 
$\alpha=0.25$. 
We denote AMR($L$,$D_2$) as the simulation for the level $L$ and the $D_2$ disk.  In particular, when $L=1$, it is referred as the uniform grid discretization. Similarly, for the $D_{2,2}$ disk, it is denoted as AMR($L$, $D_{2,2}$). 

We first demonstrate accuracy for simulations of AMR($L$, $D_2$) and AMR($L$, $D_{2,2}$), to be followed by demonstration of order of accuracy with the size  of $N_k=2^k$ zones in one dimension for $k=7,8,\ldots,11$. The results for $D_2$ and $D_{2,2}$ with $L=1$ reproduce the entries of Table~1 and Table~2 of Yen et al (2012) for $k=7$ to 10, which are shown below for comparison. Yen et al. (2012) concluded previously that the accuracy is nearly second order in the $L^1$, $L^2$ and $L^\infty$ norms. In the sense of AMR, we can express the norms as 
\begin{eqnarray*}
|u|_1 \equiv \sum_i |u_i|\Delta A_i, \quad
|u|_2 \equiv \left(\sum_i (u_i)^2\Delta A_i\right)^{1/2},\quad
|u|_\infty \equiv \max_i |u_i|
\end{eqnarray*}
where $\Delta A_i$ is the area of sub-region $D_i$. 
The order of accuracy $O_{\alpha,\beta}$ in the bottom portion of Table~1 and Table~2 are defined as $\log_2(L_{\alpha,\beta}(N_k)/L_{\alpha,\beta}(N_{k+1}))$ where
$\alpha=x,y,R$, and $\beta$ equals $1,2,\infty$. 

\begin{table}[ht]
\begin{center}
\begin{tabular}{|c|c|c|c|c|c|c|}
\hline
AMR(1,$D_2$) - $N_k$ & $L_{x,1}$ & $L_{x,2}$ & $L_{x,\infty}$ & $L_{R,1}$ & $L_{R,2}$ & $L_{R,\infty}$ \\ \hline
128 & 8.48E-4  &9.31E-4   &5.26E-3     &1.32E-3   &1.31E-3   &5.91E-3 \\ \hline
256 & 2.16E-4  &2.44E-4   &1.93E-3     &3.38E-4   &3.44E-4   &1.99E-3 \\ \hline
512 & 5.62E-5  &6.88E-5   &9.48E-4     &8.80E-5   &9.70E-5   &9.84E-4 \\ \hline
1024& 1.43E-5  &1.82E-5   &3.28E-4     &2.24E-5   &2.57E-5   &3.47E-4 \\ \hline
2048& 3.55E-6  &4.77E-6   &1.24E-4     &5.57E-6   &6.73E-6   &1.36E-4  \\ \hline
\hline
AMR(1,$D_2$) - $N_k/N_{k+1}$  &  $O_{x,1}$ & $O_{x,2}$ & $O_{x,\infty}$ & $O_{R,1}$ & $O_{R,2}$ & $O_{R,\infty}$ \\ \hline
128/256       & 1.97      &1.93      &1.45        &1.96      &1.93      &1.57 \\ \hline
256/512       & 1.94      &1.83      &1.03        &1.94      &1.83      &1.02 \\ \hline
512/1024      & 1.98      &1.92      &1.53        &1.98      &1.92      &1.50 \\ \hline
1024/2048     & 2.00      &1.93      &1.40        &2.00      &1.93      &1.35 \\ \hline
\end{tabular}
\caption{ 
AMR($L$,$D_2$) simulation for a $D_2$ disk with $L=1$ (uniform discretization). $L_{\alpha,\beta}$ represents the the errors of forces between numerical and analytic solutions
for the $x$-direction ($\alpha=x$) and radial direction ($\alpha=R$) in the norms of $\beta=1,2,\infty$. 
The top portion shows the errors for $N_k=2^k$, where $k=7,8,\ldots,11$.  The bottom portion shows the order of accuracy in the $1$- and $2$-norms are of nearly second order for both $x$ and $R$ directions, but roughly about $1.3$ for the $\infty$-norm.
}
\end{center}
\end{table}

\begin{table}[ht]
\begin{center}
\begin{tabular}{|c|c|c|c|c|c|c|c|c|c|}
\hline
AMR(1,$D_{2,2}$)-$N_k$ & $L_{x,1}$ & $L_{x,2}$ & $L_{x,\infty}$ & $L_{y,1}$ & $L_{y,2}$ & $L_{y,\infty}$ & $L_{R,1}$ & $L_{R,2}$ & $L_{R,\infty}$ \\ \hline
128 & 1.71E-3  &1.69E-3   &8.98E-3 & 1.62E-3 & 1.68E-3 & 8.98E-3    &1.72E-3   &1.71E-3   &9.06E-3 \\ \hline
256 & 4.28E-4  &4.42E-4   &3.54E-3 & 4.10E-4 & 4.42E-4 & 3.54E-3    &4.25E-4   &4.40E-4   &3.60E-3 \\ \hline
512 & 1.10E-4  &1.16E-4   &1.13E-3 & 1.05E-4 & 1.16E-4 & 1.13E-3    &1.10E-4   &1.17E-4   &1.17E-4 \\ \hline
1024& 2.84E-5  &3.13E-5   &4.78E-4 & 2.71E-5 & 3.13E-5 & 4.78E-4    &2.84E-5   &3.15E-5   &4.80E-4 \\ \hline
2048& 7.18E-6  &8.32E-6   &1.91E-4 & 6.84E-6 & 8.32E-6 & 1.91E-4    &7.18E-6   &8.36E-6   &1.81E-4 \\ \hline
\hline
AMR(1,$D_{2,2}$)-$N_k/N_{k+1}$ &  $O_{x,1}$ & $O_{x,2}$ & $O_{x,\infty}$ & $O_{y,1}$ & $O_{y,2}$ & $O_{y,\infty}$ & $O_{R,1}$ & $O_{R,2}$ & $O_{R,\infty}$ \\ \hline
128/256       & 2.00  & 1.93 & 1.34   & 1.98 & 1.93 & 1.34 & 2.02& 1.96 &  1.33\\ \hline
256/512       & 1.96  & 1.93 & 1.65   & 1.97 & 1.93 & 1.65 & 1.95& 1.91 &  1.57\\ \hline
512/1024      & 1.95  & 1.89 & 1.24   & 1.95 & 1.89 & 1.24 & 1.95& 1.89 &  1.33\\ \hline
1024/2048     & 1.98  & 1.91 & 1.32   & 1.99 & 1.91 & 1.32 & 1.98& 1.91 &  1.33\\ \hline
\end{tabular}
\caption{
Data for AMR($L$,$D_{2,2}$) simulation for $D_{2,2}$ disk with the uniform $L=1$ discretization. $L_{\alpha,\beta}$ calculates error of forces between the numerical and analytic solutions for the $x$-direction ($\alpha=x$) and the radial direction ($\alpha=R$) in $\beta=1,2,\infty$ norms, respectively. The top portion of the table shows the errors for $N_k=2^k$, where $k=7,8,\ldots,11$.  The orders of accuracy shown in the bottom portion of the table are essentially second for the $x$, $y$, and $R$ directions in the $1$- and $2$-norms, but only roughly about $1.3$ for $\infty$-norm. Such trends are the same as those shown in Table 1. 
}
\end{center}
\end{table}

\subsection{Accuracy of Computation and Order of Accuracy on AMR}

We demonstrate the cases of adaptive mesh refinement for AMR($L$,$D_2$) and AMR($L$,$D_{2,2}$) with $L=2$ and $3$. At the same time, we also investigate constraints on the refinement. We adopt two different strategies of refinement to illustrate accuracy of the method and the required time of computation under the respective conditions using two approaches of refinement. The first is the independent level refinement (ILR), where we set the distance between the edges of the next level less than 0.05 of the coarser grid for each refinement. This approach is more suitable for preserving the numerical accuracy and order of accuracy, but it may require more computational time.  The other is called Variable Level Refinement (VLR), where we set the distance between the edges of next grid level less than $0.2\times d^{1-L}$ for each of the level $L>1$ with $L=1$ being the coarsest. Here, $d=\frac{1}{2}$ is used.
The grid structures are shown in Figure 2 for $N_k=32, k=5$ and $L=3$ for the $D_2$ disk on the left and the $D_{2,2}$ disk on the right for two refinement strategies ILR on the top and VLR on the bottom.

\begin{figure}[h]
\centerline{
\includegraphics[scale=.3]{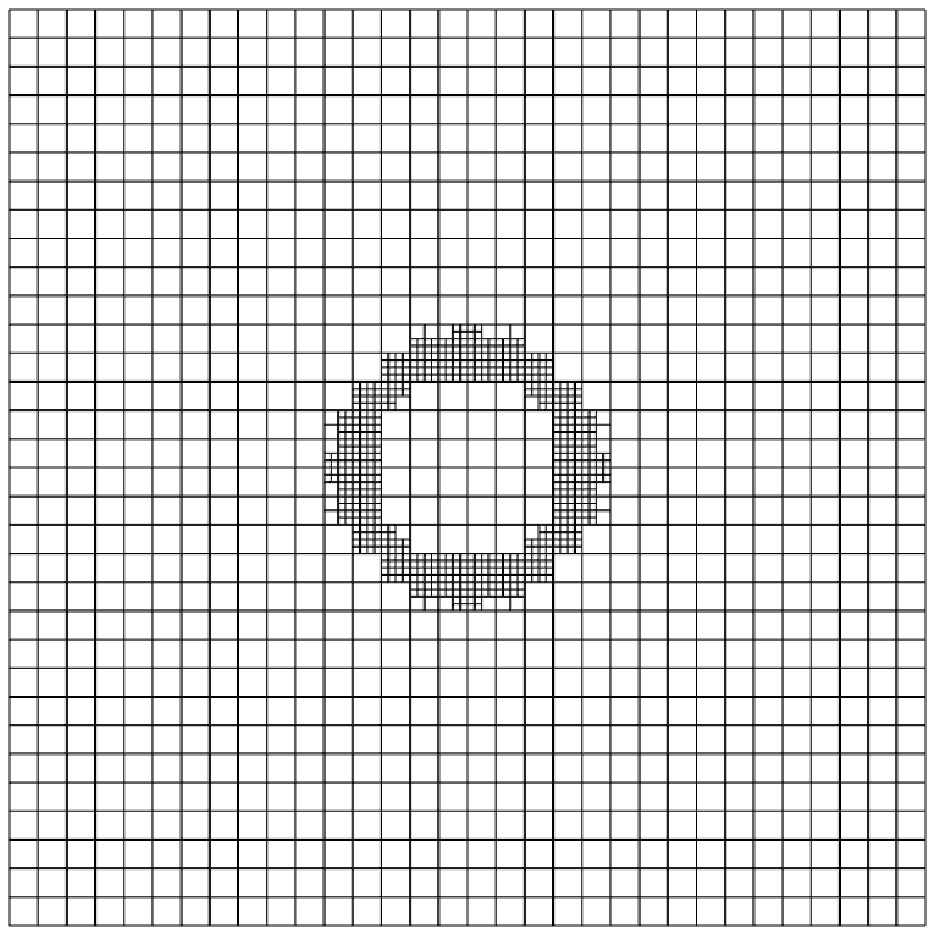}\quad 
\includegraphics[scale=.3]{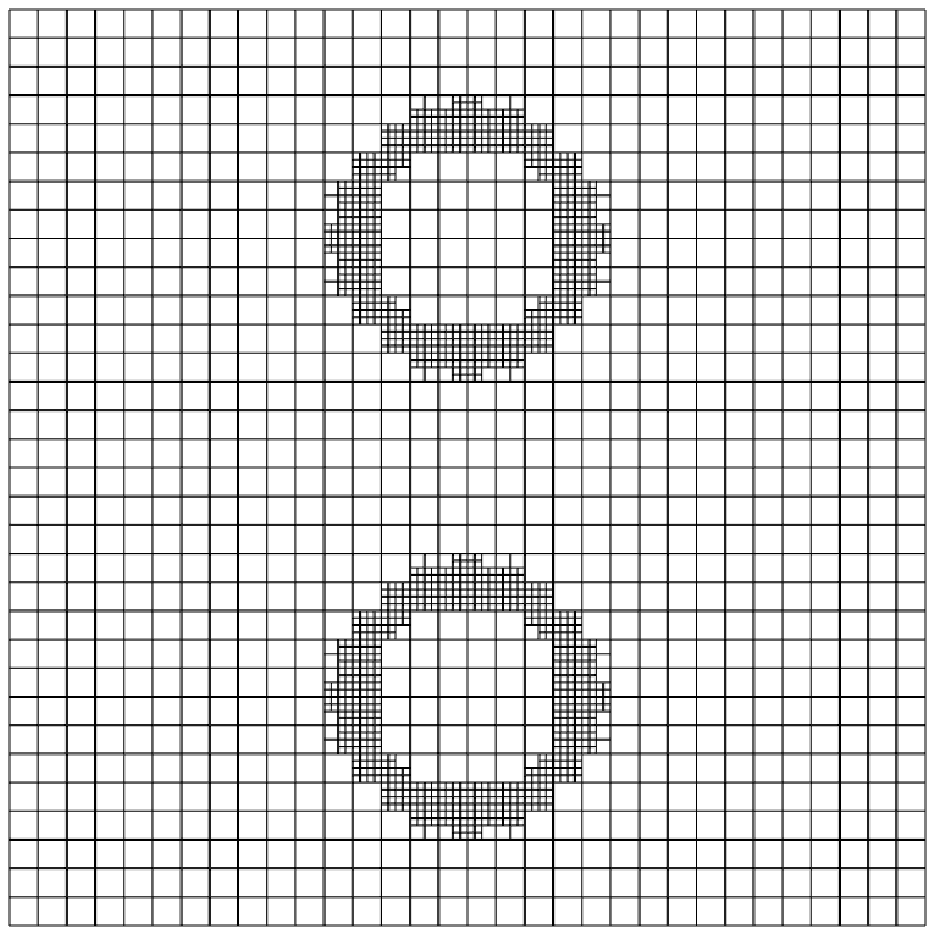}
}

\centerline{
\includegraphics[scale=.3]{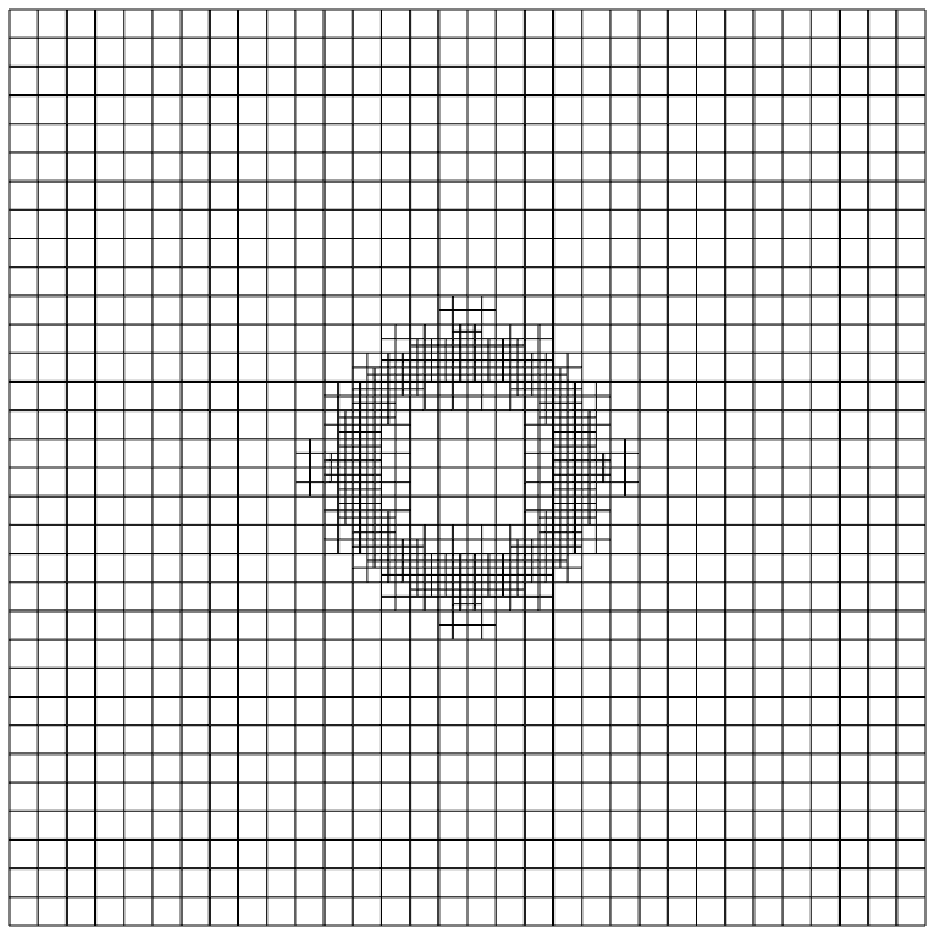}\quad 
\includegraphics[scale=.3]{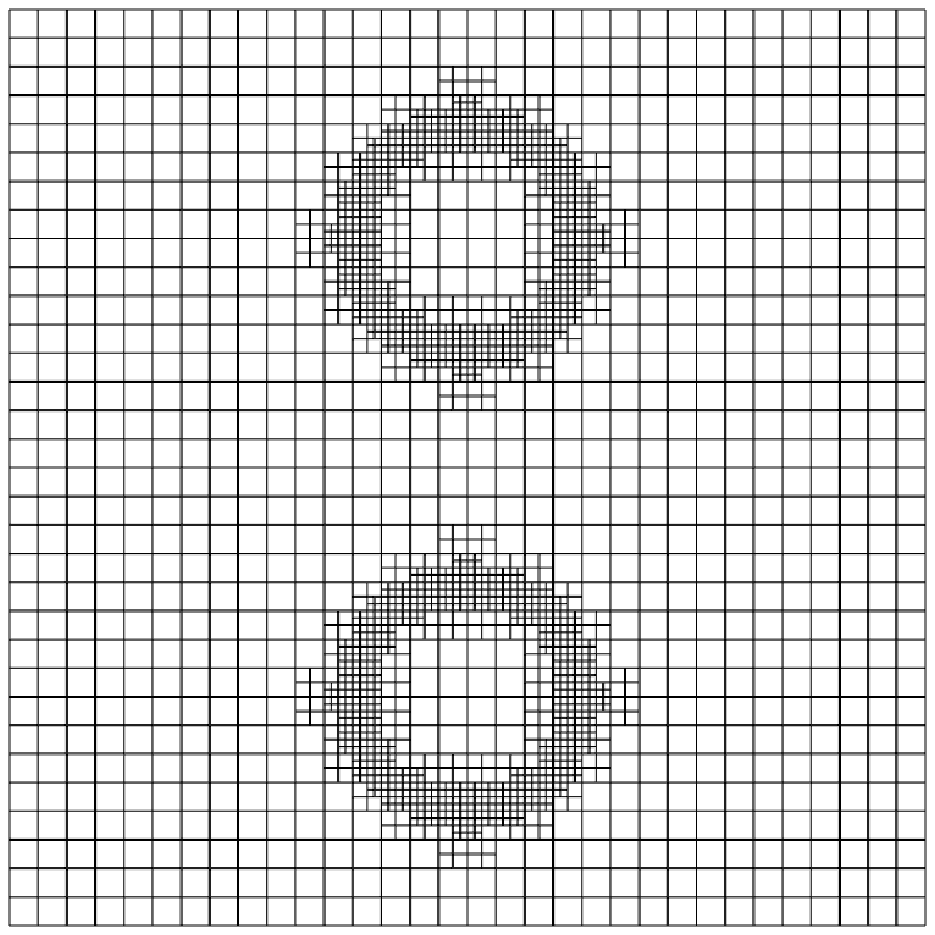}
}
\caption{Illustration of the grid structure for AMR with $N_k=2^k, k=5$ and $L=3$ for the disks $D_2$ (left) and $D_{2,2}$ (right) under the ILR refinement (top) and the VLR refinement (bottom).}
\end{figure}

The results of AMR($L$,$D_2$) for refinement levels $L=2$ and $L=3$ with the ILR are shown in Table 3. 
The first and third sub-tables are the errors
between the numerical and analytic solutions, 
$N_k=2^k$, $k=7,8,\ldots,11$, for $L=2$ and $L=3$, respectively.  
The order of accuracy is listed in the second and fourth sub-tables according to $L=2$ and $L=3$, respectively.
It is shown to be of nearly second order in both the $x$ and $R$ directions for the $1$ and $2$ norms. However, it is only about 1.4 for the $\infty$-norm.  Although the orders of accuracy are almost the same for both the uniform and AMR grids, the results obtained for the AMR converge better than those obtained for the uniform grids. This situation is shown in Figure 3. Similar results hold for the case of AMR($L$, $D_{2,2}$) for $L=2$ and $L=3$ with ILR, which are summarized in Table 4. 

\begin{table}[ht]
\begin{center}
\begin{tabular}{|c|c|c|c|c|c|c|}
\hline
AMR(2,$D_2$)-$N_k$  & $L_{x,1}$ & $L_{x,2}$ & $L_{x,\infty}$ & $L_{R,1}$ & $L_{R,2}$ & $L_{R,\infty}$ \\ \hline
$128$   & 1.67E-04  & 1.58E-04  & 1.90E-03    &  2.55E-04 & 2.22E-04  & 1.96E-03 \\ \hline
$256$   & 4.33E-05  & 4.29E-05  & 9.52E-04    &  6.71E-05 & 6.04E-05  & 9.84E-04\\ \hline
$512$   & 1.10E-05  & 1.12E-05  & 3.28E-04    &  1.72E-05 & 1.57E-05  & 3.46E-04\\ \hline
$1024$  & 2.80E-06  & 2.90E-06  & 1.24E-04    &  4.38E-06 & 4.09E-06  & 1.36E-04\\ \hline
$2048$  & 7.04E-07  & 7.53E-07  & 4.91E-05    &  1.10E-06 & 1.06E-06  & 4.93E-05\\ \hline
\hline
AMR(2,$D_2$)-$N_k/N_{k+1}$&  $O_{x,1}$ & $O_{x,2}$ & $O_{x,\infty}$ & $O_{R,1}$ & $O_{R,2}$ & $O_{R,\infty}$ \\ \hline
128/256       & 1.95  & 1.88 & 0.99  &  1.93   & 1.88    & 0.99\\ \hline
256/512       & 1.98  & 1.94 & 1.54  &  1.96   & 1.94    & 1.51\\ \hline
512/1024      & 1.97  & 1.95 & 1.40  &  1.97   & 1.94    & 1.35\\ \hline
1024/2048     & 1.99  & 1.95 & 1.34  &  1.99   & 1.95    & 1.46\\ \hline
\hline
AMR(3,$D_2$)-$N_k$  & $L_{x,1}$ & $L_{x,2}$ & $L_{x,\infty}$ & $L_{R,1}$ & $L_{R,2}$ & $L_{R,\infty}$ \\ \hline
$128$  & 1.90E-04  & 1.52E-05  & 1.14E-03    &  2.94E-04 & 2.14E-05  & 1.17E-03\\ \hline
$256$  & 4.82E-05  & 3.92E-06  & 3.62E-04    &  7.54E-05 & 5.51E-05  & 3.66E-04\\ \hline
$512$  & 1.21E-05  & 9.92E-06  & 1.30E-04    &  1.89E-05 & 1.40E-05  & 1.41E-04\\ \hline
$1024$ & 3.02E-06  & 2.52E-07  & 5.06E-05    &  4.74E-06 & 3.56E-06  & 5.09E-05\\ \hline
$2048$ & 7.57E-07  & 6.40E-07  & 1.84E-05    &  1.19E-06 & 9.04E-07  & 1.84E-05\\ \hline
\hline
AMR(3,$D_2$)-$N_k/N_{k+1}$ &  $O_{x,1}$ & $O_{x,2}$ & $O_{x,\infty}$ & $O_{R,1}$ & $O_{R,2}$ & $O_{R,\infty}$ \\ \hline
128/256    & 1.98  & 1.96 & 1.66   & 1.96     & 1.96     & 1.68 \\ \hline
256/512    & 1.99  & 1.98 & 1.48   & 2.00     & 1.98     & 1.38\\ \hline
512/1024   & 2.00  & 1.98 & 1.36   & 2.00     & 1.98     & 1.47\\ \hline
1024/2048  & 2.00  & 1.98 & 1.46   & 2.00     & 1.98     & 1.47\\ \hline
\end{tabular}
\caption{
The simulation of AMR($L$,$D_2$) for $D_2$ disk with $L=2$ and $3$ independent level refinement (ILR).
$L_{\alpha,\beta}$ represents the the errors of forces between numerical and analytic solutions
for $x$-direction ($\alpha=x$) and radial direction ($\alpha=R$) in $\beta=1,2,\infty$ norm. 
The first and third sub-tables show the errors, $N_k=2^k$, $k=7,8,\ldots,11$, for $L=2$ and $L=3$, respectively.  
The order of accuracy in the second and fourth sub-tables according to $L=2$ and $L=3$, respectively,
show that the $1$- and $2$-norms are nearly of second order for both $x$ and $R$ directions, but roughly about 
$1.4$ for $\infty$-norm. 
}
\end{center}
\end{table}

%%%%%%%%%%%%%%%%%%%%%%%%%%%%%%%%%%%%%%%%%
%%%% D_2,2
\begin{table}[ht]
\begin{center}
\begin{tabular}{|c|c|c|c|c|c|c|c|c|c|}
\hline
AMR(2,$D_{2,2}$)-$N_k$   & $L_{x,1}$ & $L_{x,2}$ & $L_{x,\infty}$ & $L_{y,1}$ &  $L_{y,2}$ & $L_{y,\infty}$  & $L_{R,1}$ & $L_{R,2}$ & $L_{R,\infty}$ \\ \hline
$128$   &3.50E-04& 5.60E-04 & 3.46E-03  & 3.12E-04  & 5.50E-04  & 3.57E-03 & 3.53E-04 & 5.58E-04 & 3.51E-03\\ \hline
$256$   &8.96E-05& 1.45E-04 & 1.09E-03  & 8.02E-05  & 1.43E-04  & 1.11E-03 & 9.08E-05 & 1.45E-04 & 1.16E-03\\ \hline
$512$   &2.30E-05& 3.82E-05 & 4.69E-04  & 2.07E-05  & 3.76E-05  & 4.76E-04 & 2.34E-05 & 3.82E-05 & 4.70E-04\\ \hline
$1024$  &5.86E-06& 9.99E-06 & 1.89E-04  & 5.29E-06  & 9.86E-06  & 1.91E-04 & 5.95E-06 & 1.00E-05 & 1.89E-04\\ \hline
$2048$  &1.47E-06& 2.58E-06 & 6.73E-05  & 1.33E-06  & 2.56E-06  & 6.78E-05 & 1.48E-06 & 2.58E-06 & 6.76E-05\\ \hline
\hline
AMR(2,$D_{2,2}$)-$N_k/N_{k+1}$   & $O_{x,1}$ & $O_{x,2}$ & $O_{x,\infty}$ & $O_{y,1}$ &  $O_{y,2}$ & $O_{y,\infty}$  & $O_{R,1}$ & $O_{R,2}$ & $O_{R,\infty}$ \\ \hline
$128/256$   & 1.96  &1.95 & 1.67  & 1.96 & 1.94  & 1.69 & 1.96 & 1.94 & 1.60 \\ \hline
$256/512$   & 1.96  &1.92 & 1.22  & 1.95 & 1.93  & 1.22 & 1.96 & 1.92 & 1.30\\ \hline
$512/1024$  & 1.97  &1.94 & 1.31  & 1.97 & 1.93  & 1.32 & 1.98 & 1.93 & 1.31\\ \hline
$1024/2048$ & 2.00  &1.95 & 1.49  & 1.99 & 1.95  & 1.49 & 2.01 & 1.95 & 1.48\\ \hline
\hline
AMR(3,$D_{2,2}$)-$N_k$   & $L_{x,1}$ & $L_{x,2}$ & $L_{x,\infty}$ & $L_{y,1}$ &  $L_{y,2}$ & $L_{y,\infty}$  & $L_{R,1}$ & $L_{R,2}$ & $L_{R,\infty}$ \\ \hline
$128$   &8.72E-04&2.50E-04 & 1.78E-03  & 7.76E-05  & 2.44E-04  & 1.79E-03 & 9.14E-05 & 2.52E-04 & 1.79E-03\\ \hline
$256$   &2.17E-05&6.36E-05 & 4.66E-04  & 1.93E-05  & 6.23E-05  & 5.04E-04 & 2.28E-05 & 6.40E-05 & 4.77E-04\\ \hline
$512$   &5.52E-06&1.63E-05 & 1.87E-04  & 4.93E-06  & 1.60E-05  & 1.97E-04 & 5.80E-06 & 1.64E-05 & 1.95E-04\\ \hline
$1024$  &1.39E-06&4.14E-06 & 6.70E-05  & 1.24E-06  & 4.06E-06  & 6.94E-05 & 1.46E-06 & 4.15E-06 & 6.72E-05\\ \hline
$2048$  &3.50E-07&1.05E-06 & 2.46E-05  & 3.15E-07  & 1.04E-06  & 2.51E-05 & 3.65E-07 & 1.06E-06 & 2.46E-05 \\ \hline
\hline
AMR(3,$D_{2,2}$)-$N_k/N_{k+1}$   & $O_{x,1}$ & $O_{x,2}$ & $O_{x,\infty}$ & $O_{y,1}$ &  $O_{y,2}$ & $O_{y,\infty}$  & $O_{R,1}$ & $O_{R,2}$ & $O_{R,\infty}$ \\ \hline
$128/256$   &2.01    & 1.97& 1.94 & 2.01 & 1.97 & 1.83 & 2.00 & 1.98 & 1.91\\ \hline
$256/512$   &1.98    & 1.96& 1.32 & 1.97 & 1.96 & 1.36 & 1.97 & 1.96 & 1.29\\ \hline
$512/1024$  &1.99    & 1.98& 1.48 & 1.99 & 1.98 & 1.50 & 1.99 & 1.97 & 1.54\\ \hline
$1024/2048$ &1.99    & 1.98& 1.45 & 1.98 & 1.96 & 1.47 & 2.00 & 1.97 & 1.45\\ \hline

\end{tabular}
\caption{
AMR($L$,$D_{2,2}$) simulation for $D_{2,2}$ disk with $L=2$ and $3$ independent level refinement (ILR). 
$L_{\alpha,\beta}$ represents the the errors of forces between numerical and analytic solutions
for $x$-direction ($\alpha=x$) and radial direction ($\alpha=R$) in $\beta=1,2,\infty$ norm. 
The first and third sub-tables show the errors, $N_k=2^k$, $k=7,8,\ldots,11$, for $L=2$ and $L=3$, respectively.  
The order of accuracy in the second and fourth sub-tables according to $L=2$ and $L=3$, respectively,
show that the $1$- and $2$-norm are nearly of second order for both $x$ and $R$ directions, but roughly about $1.4$ for $\infty$-norm. 
}
\end{center}
\end{table}

We also investigate the Variable Level Refinement approach for the $D_2$ disk and $D_{2,2}$ disk. The results of AMR($L$,$D_2$) for $L=2$ and $L=3$ with VLR are shown in Table 5.  The first and third sub-tables show the errors, $N_k=2^k$, $k=7,8,\ldots,11$, for $L=2$ and $L=3$, respectively.  The order of accuracy is listed in the second and fourth sub-tables according to $L=2$ and $L=3$, respectively.
From the Table 5, the values for the order of accuracy fall between 0.8 and 3 among the $1$, $2$, $\infty$-norms.  
This shows that the convergence can go beyond the second order, which is the phenomenon of super-convergence. Although the order of accuracy fluctuates more for the AMR grid with VLR, the majority of results with the AMR are better converged than some of the uniform grids. Similar situation holds for the AMR($L$, $D_{2,2}$) simulation for $L=2$ and $L=3$ with VLR in Table 6, but with a narrower range of order of accuracy than those of AMR($L$,$D_2$).
%%%%%%

\begin{table}[ht]
\begin{center}
\begin{tabular}{|c|c|c|c|c|c|c|}
\hline
AMR(2,$D_2$)-$N_k$  & $L_{x,1}$ & $L_{x,2}$ & $L_{x,\infty}$ & $L_{R,1}$ & $L_{R,2}$ & $L_{R,\infty}$ \\ \hline
$128$   & 2.70E-04  & 1.91E-04  & 1.65E-03    &  4.22E-04 & 1.70E-04  & 1.71E-03 \\ \hline
$256$   & 5.05E-05  & 4.26E-05  & 8.92E-04    &  7.86E-05 & 6.00E-05  & 9.29E-04\\ \hline
$512$   & 1.10E-05  & 1.12E-05  & 3.28E-04    &  1.72E-05 & 1.57E-05  & 3.46E-04\\ \hline
$1024$  & 2.91E-06  & 3.28E-06  & 1.30E-04    &  4.56E-06 & 4.62E-06  & 1.42E-04\\ \hline
$2048$  & 7.89E-07  & 9.62E-07  & 5.32E-05    &  1.24E-06 & 1.36E-06  & 5.34E-05\\ \hline
\hline
AMR(2,$D_2$)-$N_k/N_{k+1}$&  $O_{x,1}$ & $O_{x,2}$ & $O_{x,\infty}$ & $O_{R,1}$ & $O_{R,2}$ & $O_{R,\infty}$ \\ \hline
128/256       &  1.16 & 0.89 & 2.42   & 2.42     &  2.17    & 0.88 \\ \hline
256/512       &  1.93 & 1.44 & 2.20   & 2.20     &  1.93    & 1.42\\ \hline
512/1024      &  1.77 & 1.33 & 1.92   & 1.92     &  1.76    & 1.28\\ \hline
1024/2048     &  1.77 & 1.29 & 1.88   & 1.88     &  1.76    & 1.41\\ \hline
\hline
AMR(3,$D_2$)-$N_k$  & $L_{x,1}$ & $L_{x,2}$ & $L_{x,\infty}$ & $L_{R,1}$ & $L_{R,2}$ & $L_{R,\infty}$ \\ \hline
$128$   & 1.15E-04  & 5.14E-05  & 1.01E-03    &  1.80E-04 & 7.27E-05  & 1.15E-03 \\ \hline
$256$   & 1.44E-05  & 9.10E-06  & 2.72E-05    &  2.25E-05 & 1.28E-05  & 2.91E-04\\ \hline
$512$   & 3.01E-06  & 2.48E-06  & 1.30E-04    &  4.72E-06 & 3.50E-06  & 1.42E-04\\ \hline
$1024$ & 9.82E-07  & 8.66E-07  & 5.93E-05    &  1.54E-06 & 1.22E-06  & 5.96E-05\\ \hline
$2048$ & 2.63E-07  & 2.80E-07  & 2.39E-05    &  4.17E-07 & 3.95E-07  & 2.39E-05\\ \hline
\hline
AMR(3,$D_2$)-$N_k/N_{k+1}$ &  $O_{x,1}$ & $O_{x,2}$ & $O_{x,\infty}$ & $O_{R,1}$ & $O_{R,2}$ & $O_{R,\infty}$ \\ \hline
128/256       &  2.50 & 1.90 & 3.00   & 3.00     &  2.51    & 1.98 \\ \hline
256/512       &  1.88 & 1.07 & 2.26   & 2.25     &  1.87    & 1.05\\ \hline
512/1024      &  1.52 & 1.13 & 1.62   & 1.62     &  1.52    & 1.24\\ \hline
1024/2048     &  1.63 & 1.31 & 1.90   & 1.90     &  1.63    & 1.32\\ \hline
\end{tabular}
\caption{
AMR($L$,$D_2$) simulation for a $D_2$ disk with $L=2$ and $3$ refinements for variable levels of refinement (VLR).
$L_{\alpha,\beta}$ represents the the errors of forces between numerical and analytic solutions
for $x$-direction ($\alpha=x$) and radial direction ($\alpha=R$) in $\beta=1,2,\infty$ norm. 
The first and third sub-tables show the errors, $N_k=2^k$, $k=7,8,\ldots,11$, for $L=2$ and $L=3$, respectively.  
The order of accuracy in the second and fourth sub-tables according to $L=2$ and $L=3$, respectively, dense in $N_k$ for each norm, $1$, $2$, $\infty$-norm between 0.88 and 3.
}
\end{center}
\end{table}

%%%%%%%%%%%%%%%%%%%%%%%%%%%%%%%%%%%%%%%%%
%%%% D_2,2
\begin{table}[ht]
\begin{center}
\begin{tabular}{|c|c|c|c|c|c|c|c|c|c|}
\hline
AMR(2,$D_{2,2}$)-$N_k$   & $L_{x,1}$ & $L_{x,2}$ & $L_{x,\infty}$ & $L_{y,1}$ &  $L_{y,2}$ & $L_{y,\infty}$  & $L_{R,1}$ & $L_{R,2}$ & $L_{R,\infty}$ \\ \hline
$128$  &6.20E-04&7.31E-04 & 3.31E-03  & 5.42E-04  & 7.03E-04 & 3.26E-03 & 6.28E-04 & 7.28E-04 & 3.15E-03\\ \hline
$256$  &1.13E-04&1.55E-04 & 1.03E-03  & 9.84E-05  & 1.49E-04 & 1.06E-03 & 1.15E-04 & 1.55E-04 & 1.10E-03\\ \hline
$512$  &2.30E-05&3.82E-05 & 4.69E-04  & 2.07E-05  & 3.76E-05 & 4.76E-04 & 2.34E-05 & 3.82E-05 & 4.70E-04\\ \hline
$1024$ &5.71E-06&1.10E-05 & 1.97E-04  & 5.41E-06  & 1.10E-05 & 1.98E-04 & 5.77E-06 & 1.10E-05 & 1.97E-04\\ \hline
$2048$ &1.56E-06&3.25E-06 & 7.28E-05  & 1.52E-06  & 3.24E-06 & 7.29E-05 & 1.56E-06 & 3.25E-06 & 7.32E-05\\ \hline
\hline
AMR(2,$D_{2,2}$)-$N_k/N_{k+1}$   & $O_{x,1}$ & $O_{x,2}$ & $O_{x,\infty}$ & $O_{y,1}$ &  $O_{y,2}$ & $O_{y,\infty}$  & $O_{R,1}$ & $O_{R,2}$ & $O_{R,\infty}$ \\ \hline
$128/256$   & 2.46   & 2.24 &  1.59 & 2.46 & 2.24  &  1.62 & 2.45 & 2.23 & 1.52 \\ \hline
$256/512$   & 2.30   & 2.02 &  1.14 & 2.25 & 1.99  &  1.16 & 2.30 & 2.02 & 1.23\\ \hline
$512/1024$  & 2.01   & 1.80 &  1.25 & 1.94 & 1.77  &  1.27 & 2.02 & 1.80 & 1.25\\ \hline
$1024/2048$ & 1.87   & 1.76 &  1.44 & 1.83 & 1.76  &  1.44 & 1.89 & 1.76 & 1.43 \\ \hline
\hline
AMR(3,$D_{2,2}$)-$N_k$   & $L_{x,1}$ & $L_{x,2}$ & $L_{x,\infty}$ & $L_{y,1}$ &  $L_{y,2}$ & $L_{y,\infty}$  & $L_{R,1}$ & $L_{R,2}$ & $L_{R,\infty}$ \\ \hline
$128$  &2.87E-04&4.14E-04 & 1.73E-03  & 2.36E-04  & 3.75E-04 & 1.75E-03 & 2.91E-04 & 4.14E-04 & 1.86E-03\\ \hline
$256$  &3.87E-05&7.30E-05 & 4.15E-04  & 3.08E-05  & 6.66E-05 & 4.20E-04 & 3.95E-05 & 7.32E-05 & 4.20E-04\\ \hline
$512$  &5.52E-06&1.63E-05 & 1.87E-04  & 4.93E-06  & 1.60E-05 & 1.97E-04 & 5.80E-06 & 1.64E-05 & 1.95E-04\\ \hline
$1024$ &1.88E-06&5.53E-06 & 7.89E-05  & 1.80E-06  & 5.55E-06 & 8.03E-05 & 1.90E-06 & 5.54E-06 & 7.91E-05\\ \hline
$2048$ &5.26E-07&1.85E-06 & 3.21E-05  & 5.14E-07  & 1.85E-06 & 3.23E-05 & 5.27E-06 & 1.85E-06 & 3.21E-05\\ \hline
\hline
AMR(3,$D_{2,2}$)-$N_k/N_{k+1}$   & $O_{x,1}$ & $O_{x,2}$ & $O_{x,\infty}$ & $O_{y,1}$ &  $O_{y,2}$ & $O_{y,\infty}$  & $O_{R,1}$ & $O_{R,2}$ & $O_{R,\infty}$ \\ \hline
$128/256$   & 2.89   & 2.50& 2.06 & 2.94 & 2.49  & 2.06 & 2.88 & 2.50 & 2.15 \\ \hline
$256/512$   & 2.81   & 2.16& 1.15 & 2.64 & 2.06  & 1.09 & 2.77 & 2.16 & 1.11\\ \hline
$512/1024$  & 1.55   & 1.56& 1.24 & 1.45 & 1.53  & 1.29 & 1.61 & 1.57 & 1.30\\ \hline
$1024/2048$ & 1.84   & 1.58& 1.30 & 1.81 & 1.59  & 1.31 & 1.85 & 1.58 & 1.30\\ \hline
\end{tabular}
\caption{
AMR($L$,$D_{2,2}$) simulation for $D_{2,2}$ disk with $L=2$ and $3$ with various level refinement (VLR).
$L_{\alpha,\beta}$ represents the the errors of forces between numerical and analytic solutions
for $x$-direction ($\alpha=x$) and radial direction ($\alpha=R$) in $\beta=1,2,\infty$ norm. 
The first and third sub-tables show the errors, $N_k=2^k$, $k=7,8,\ldots,11$, for $L=2$ and $L=3$, respectively.  
The order of accuracy in the second and fourth sub-tables according to $L=2$ and $L=3$, respectively, dense in $N_k$ for each norm, $1$, $2$, $\infty$-norm between 1.11 and 2.94.
The result is the same as Table 5.
}
\end{center}
\end{table}

\subsection{Comparison of Run-Time on GPUs and on CPUs}

The direct $N$-body method puts a strict demand on the computing power to deal with the calculation of self-gravitational force. Parallel computing offers a great advantage in term of performance because it deploys the same calculation to lots of processing units to run in the same amount of time. 
MPI and OpenMP are popular parallel schemes for the standard computers.

Until about a decade ago did the potential of using graphical processing units (GPU) for general purpose applications become viable. In 2006, Nvidia Corporation introduced a general purpose parallel programming architecture that uses their parallel computing engine GPUs. The platform, Compute Unified Device Architecture, abbreviated as CUDA, is a software layer that gives direct access to the virtual instruction set of Nvidia GPUs and parallel computational elements. In addition, it solves complex computational problems in a more efficient way with much better performance than a CPU does, especially on direct $N$-body calculation like the one illustrated in this paper. 
OpenACC is another different parallel programming model which is user-driven, and directives-based on heterogeneous hardware platforms and architectures. Just like OpenMP, user can add few directives into the source code and specify how a compiler should process these parts with GPU.   
The OpenACC specifications supports Fortran, C, C++ programming language and a variety of different architectures including AMD HD GPUs, Nvidia  GPUs and Intel Xeon Phi. Although the performance of OpenACC is not good as CUDA, OpenACC is easier to implement
into an existing code.

In this paper, we employ both Fortran and CUDA languages in the simulations for comparison of performance. The GPU computations were run entirely on Nvidia Tesla P100 with CUDA. Tesla P100 is powered by the Pascal GPU microarchitecture and was released in 2016. Each Tesla P100 has 3584 Nvidia CUDA cores with a GPU clock speed of 1189 MHz. The double-precision performance is 4.7 teraFLOPS according to the Nvidia website. The Fortran CPU code runs on Intel Xeon E5-2698 operating at 2.2GHz. The Xeon E5-2698 is a 64-bit icosa-core x86 microprocessor introduced by Intel in 2016. Although Xeon E5-2698 has 20 cores, each Fortran benchmark runs on a single core.

A direct method as illustrated here is a time intensive one. It also requires investigations on how realistically capabilities of the methodology can be carried with available computational resources. For this purpose, we compare simulations carried out with the Fortran and CUDA platforms. The AMR($1$,$D_2$) and AMR($1$,$D_{2,2}$) were carried out with both Fortran and CUDA platforms for comparison. Table 7 shows that the CUDA platform is faster than the Fortran one by about 263 times, i.e, CT(Fortran)=263 $\times$ CT(CUDA), where CT stands for the computational time. This can be seen directly in Table 7, the column of Fortran/CUDA at $N_k=512$. Therefore, GPU can rise to the task for simulations in practice and we compare the performance solely in the simulations with AMR grids for the two refinement strategies, ILR and VLR, as below. Tables 8 and 9 summarize computational times for AMR($L$,$D_2$) and AMR($L$,$D_{2,2}$) based only on the CUDA platform. On average, the computational time for the AMR, CT(AMR), is about 8 times of the computational time for the uniform grids (UG, $L=1$), CT(UG), from the column AMR/UG. The complexity of computational times for self-gravity is related to the number of leaf zones. In Table 8, the $EL/(N^2_k)$ ratios for $N_k=2^k$, $k=7,8,\ldots,11$ and $D_2$ using ILR are fixed at $1.12$ and $1.36$ for $L=2$ and $L=3$, respectively. The $EL$ is the number of the leaf zones which are interacting with each other by gravity for the AMR. The $N^2_k$ is the zone number of the uniform grid. The higher $EL/(N^2_k)$ ratio, the more gravity calculation for the AMR. Likewise for $D_{2,2}$, the ratios are $1.19$ and $1.57$ for $L=2$ and $L=3$, respectively. 
The computational time increases exponentially with $N_k$. The behavior is similar for VLR as reported in Table 9.

\begin{table}
\begin{center}
%\begin{tabular}{|c|c|c|c|}
\begin{tabular}{|c|c|c|c|}
\hline
%AMR(1, $D_2$)-$N_k$& Fortran (sec) & Open ACC (sec) & CUDA (sec)\\ \hline
%128& 1.58E+2& 4.90E+0 & 4.29E+0\\ \hline
%%256& 2.71E+3& 7.12E+0 & 7.28E+0\\ \hline
%512& 4.11E+4& 4.31E+1 & 1.56E+1 \\ \hline
%1024 &    $-$       & 6.03E+2 & 1.71E+2\\ \hline
%2048 &    $-$       & 9.49E+3  & 2.63E+3\\ \hline
%\hline
%AMR(1, $D_{2,2}$)-$N_k$& Fortran (sec) & Open ACC (sec) & CUDA (sec)\\ \hline
%128& 1.62E+2 &  & 4.39E+0\\ \hline
%256& 2.68E+3 &  & 5.29E+0\\ \hline
%512&  $-$    &  & 1.57E+1\\ \hline
%1024&    $-$&  & 1.71E+2\\ \hline
%2048&    $-$&  & 2.62E+3\\ \hline
AMR(1, $D_2$)-$N_k$& Fortran (sec)  & CUDA (sec) & Fortran/CUDA\\ \hline
$128$& 1.58E+2&  4.29E+0 & 3.68E+1\\ \hline
$256$& 2.71E+3&  7.28E+0 & 3.72E+2\\ \hline
$512$& 4.11E+4&  1.56E+1 & 2.63E+2\\ \hline
$1024$ &    $-$  & 1.71E+2 & *\\ \hline
$2048$ &    $-$  & 2.63E+3 & *\\ \hline
\hline
AMR(1, $D_{2,2}$)-$N_k\times N_k$& Fortran (sec)& CUDA (sec) & Fortran/CUDA\\ \hline
$128\times 128$& 1.62E+2 &   4.39E+0 & 3.69E+1\\ \hline
$256\times 256$& 2.68E+3 &   5.29E+0 & 5.07E+2\\ \hline
$512\times 512$&  $-$    &   1.57E+1 & *\\ \hline
$1024\times 1024$& $-$    &   1.71E+2 &* \\ \hline
$2048\times 2048$& $-$    &   2.62E+3 &* \\ \hline
\end{tabular}
\end{center}
\caption{
The computational time for AMR(1,$D_2$) (top) and AMR(1,$D_{2,2}$) (bottom) simulation by Fortran and  CUDA. 
The symbol ``$-$'' means that the computational time is more than one week and ``*'' means that no calculation. The result shows that CUDA is faster than Fortran about 263 times at $N_k=512$.
}
\end{table}

\begin{table}[ht]
\begin{center}
\begin{tabular}{|c|c|c|c|c|c|c|c|}
\hline
AMR(2, $D_2$)-$N_k$   & $ L=1; N^2_k$ &  $L=2$   & $EL$(leaf)  & $EL/(N^2_k)$ & CT(sec) & CT($N_{k}$)/CT($N_{k-1}$) &  AMR/UG\\ \hline
$128$      & 16384    &  19056     & 18388       &  1.12       &                4.55E+0  &$-$& 1.06\\ \hline
$256$      & 65536    &  75792     & 73228       &  1.12       &                7.28E+0  &1.60& 1.00\\ \hline
$512$      & 262144   &  303344    & 293044      &  1.12       &                4.65E+1  &6.39& 2.98\\ \hline
$1024$    & 1048576  &  1213376   & 1172176     &  1.12       &                6.71E+2  &14.43& 3.93\\ \hline
$2048$    & 4194304  &  4853408   & 4688632     &  1.12       &                1.06E+4  &15.80& 4.03\\ \hline
\hline
AMR(3, $D_2$)-$N_k$   & $L=1; N^2_k$  &   $L=3$   & $EL$ (leaf)  &  $EL/N^2_k$ &   CT(sec) &
CT($N_{k}$)/CT($N_{k-1}$) & AMR/UG \\ \hline
$128$       & 16384    &   24272   & 22300       & 1.36       &              4.62E+0   &$-$& 1.08\\ \hline
$256$       & 65536    &   96256   & 88576       & 1.35       &              8.67E+0   &1.88& 1.19\\ \hline
$512$       & 262144   &   385712  & 354820      & 1.35       &              7.01E+1   &8.09& 4.49\\ \hline
$1024$     & 1048576  &   1543040 & 1419424     & 1.35       &              1.05E+3   &14.98& 6.14\\ \hline
$2048$     & 4194304  &   6171680 & 5677336     & 1.35       &              1.62E+4   &15.43& 6.16\\ \hline
\hline
AMR(2, $D_{2,2}$)-$N_k$  & $ L=1; N^2_k$ &  $L=2$ & $EL$(leaf)  & $EL/(N^2_k)$ & CT(sec) & 
CT($N_{k}$)/CT($N_{k-1}$) & AMR/UG \\ \hline
$128$                 & 16384         &  20544    & 19504    & 1.19       &  4.59E+0 &$-$& 1.05\\ \hline
$256$                 & 65536         &  81984    & 77872    & 1.19       &  8.19E+0 &1.78& 1.55\\ \hline
$512$                 & 262144        &  327904   & 311464   & 1.19       &  6.71E+1 &8.19& 4.27\\ \hline
$1024$               & 1048576       &  1312192  & 1246288  & 1.19       &  1.07E+3 &15.99& 6.26\\ \hline
$2048$               & 4194304       &  5248128  & 4984672  & 1.19       &  1.28E+4 &11.88& 4.89\\ \hline
\hline
AMR(3, $D_{2,2}$)-$N_k$   & $L=1; N^2_k$ &   $L=3$   & $EL$ (leaf)  &  $EL/N^2_k$ &  CT (sec) & 
CT($N_{k}$)/CT($N_{k-1}$) & AMR/UG \\ \hline
$128$                 & 16384         & 28800     & 25696      & 1.57   & 4.87E+0  &$-$& 1.11\\ \hline
$256$                 & 65536         & 114848    & 102520     & 1.56   & 1.19E+1  &2.44& 2.25\\ \hline
$512$                 & 262144        & 459648    & 410272     & 1.57   & 1.27E+2  &10.67& 8.09\\ \hline
$1024$               & 1048576       & 1838720   & 1641184    & 1.57    & 2.01E+3  &15.83& 11.75\\ \hline
$2048$               & 4194304       & 7356544   & 6565984    & 1.57    & 2.25E+4  &11.19& 8.59\\ \hline
\end{tabular}
\caption{
The computational time for AMR($L$,$D_2$)
and AMR($L$, $D_{2,2}$) for the level $L=2$ and $3$ simulations with ILR by CUDA. 
The $N_k$ is the number of zones in one dimension. 
The numbers in $L=2$ or $L=3$ are the total numbers of zones,
$EL$(leaf) is the corresponding number of leaf zones. 
CT means the computational time. 
The $EL/N^2_k$ is the ratio of the number of leaf zones divided by $N^2_k$. 
The ratios CT($N_k$)/CT($N_{k-1}$) show that the computational time exponentially grows with $N_k$
for $k=8,9,10,11$
}
\end{center}
\end{table}

\begin{table}[ht]
\begin{center}
\begin{tabular}{|c|c|c|c|c|c|c|c|}
\hline
AMR(2, $D_2$)-$N_k$   & $ L=1; N^2_k$ &  $L=2$     & $EL$(leaf)  & $EL/(N^2_k)$   & CT(sec) &
CT($N_{k}$)/CT($N_{k-1}$) & AMR/UG \\ \hline
$128$   & 16384                 &  26672     & 24100 &  1.47            &4.61E+0 &$-$& 1.07\\ \hline
$256$   & 65536                 &  86032     & 80908 &  1.23            &8.08E+0 &1.75& 1.11\\ \hline
$512$   & 262144                & 303344     & 293044 & 1.18            &4.64E+1 &5.74& 2.97\\ \hline
$1024$ & 1048576               & 1130944    & 1110352& 1.06            &5.99E+2 &12.91& 3.50\\ \hline
$2048$ & 4194304               & 4358832    & 4317700& 1.03            &9.02E+3 &15.06& 3.43\\ \hline
\hline
AMR(3, $D_2$)-$N_k$   & $L=1; N^2_k$ &   $L=3$   & $EL$ (leaf) &  $EL/N^2_k$     & CT(sec)&
CT($N_{k}$)/CT($N_{k-1}$) & AMR/UG\\ \hline
$128$   & 16384       &  47168    & 39472        &   2.41      &   5.36E+0 &$-$& 1.25\\ \hline
$256$   & 65536       &   127232 &111808         &   1.71      &   1.25E+1 &2.33& 1.72\\ \hline
$512$   & 262144      &   385712  &354820        &   1.35      &   6.98E+1 &5.58& 4.47\\ \hline
$1024$ & 1048576     &   1295472 &1233748       &   1.18      &   7.94E+2 &11.37& 4.64\\ \hline
$2048$ & 4194304     &   4688352 &4564840       &   1.09      &   1.09E+4 &13.73& 4.14\\ \hline
\hline
AMR(2, $D_{2,2}$)-$N_k$   & $ L=1; N^2_k$ &  $L=2$     & $EL$(leaf)  & $EL/(N^2_k)$ &  CT(sec) & 
CT($N_{k}$)/CT($N_{k-1}$) & AMR/UG\\ \hline
$128$   & 16384           &  32928     & 28792  & 1.76 &           4.97E+0 &$-$& 1.13\\ \hline
$256$   & 65536           &  98400     & 90184  & 1.37 &           9.25E+0 &1.86& 1.75\\ \hline
$512$   & 262144          & 327904     & 311464 & 1.19 &           7.03E+1 &7.60& 4.48\\ \hline
$1024$ & 1048576         & 1180320    & 1147384& 1.09 &           2.01E+3 &28.59& 11.75\\ \hline
$2048$ & 4194304         & 4458080    & 4392136& 1.05 &           8.45E+4 &42.04& 32.13\\ \hline
\hline
AMR(3, $D_{2,2}$)-$N_k$   & $L=1; N^2_k$ &   $L=3$   & $EL$ (leaf)  &  $EL/N^2_k$ &   CT(sec) & 
CT($N_{k}$)/CT($N_{k-1}$) & AMR/UG\\ \hline
$128$   & 16384        &  65792    & 53440       & 3.26     &      7.07E+0 &$-$& 1.61\\ \hline
$256$   & 65536        &  164160   &139504       & 2.13     &      1.59E+1 &2.25& 3.01\\ \hline
$512$   & 262144       &  459648   &410272       & 1.57     &      1.27E+2 &7.99& 8.09\\ \hline
$1024$ & 1048576      &  1444096  &1345216      & 1.28     &      2.53E+3 &19.92& 14.80\\ \hline
$2048$ & 4194304      &  4985152  &4787440      & 1.14     &      1.28E+5 &50.59& 48.86\\ \hline
\end{tabular}
\caption{
The computational time for AMR($L$,$D_2$)
and AMR($L$, $D_{2,2}$) for the level $L=2$ and $3$ simulations with VLR by CUDA. $N_k$ is number of zones in one dimension. 
The numbers in $L=2$ or $L=3$ are the total numbers of zones, $EL$(leaf) is the corresponding number of leaf zones. CT stands for the computational time. The $EL/N^2_k$ is the ratio of the number of leaf zones divided by $N^2_k$. The ratios CT($N_k$)/CT($N_{k-1}$) shows that the computational time exponentially grows with $N_k$ for $k=8,9,10,11$
}
\end{center}
\end{table}

\section{Discussion and Conclusion}
\label{sec:Discussion and Conclusion}

We demonstrated a direct method of second order accuracy for self-gravitational force calculation based on AMR by numerical simulations. The AMR grids can further improve the accuracy but preserving the order of accuracy, as shown in Figures 3 and 4 for the disk $D_2$ and the disk $D_{2,2}$, respectively. 
The values of errors are from Table 1 and Table 2 for uniform grids, from Table 3 and Table 4 for AMR with ILR, and from Table 5 and Table 6 for AMR with VLR. 
The slopes of the lines on the $log_2$(error)-$log_2(N_k)$ plots give the order of accuracy while the absolute $log_2$(error) position give the accuracy of the computation. These profiles in Figures 3 and 4 demonstrate that the $L=3$ refinement indeed gives the most accurate calculations among the simulations in this work for $D_2$ and $D_{2,2}$ disks, and likewise, the $L=2$ refinement is better than the $L=1$ uniform grid. Profiles of errors in the $x$, $y$ and radial directions for the $L=3$ AMR($3$, $D_{2,2}$) with ILR using $N_k=2048$ are shown in Figure 5. It illustrates that most errors occur near the edges of the disks.

Comparing the achieved accuracy and computational time with $N_k$, the ILR approach behaves well in terms of the order of accuracy. That is, the slopes of lines for ILR simulations in Figure 3 are almost identical for the levels of $L=1,2,3$. On the other hand, the computational times of VLR are always less than those of the ILR at $N_k=512$ for $L=2,3$.  Although the order of accuracy for VLR fluctuates between 0.88 and 3 for the $D_2$ disk and between 1.11 and 2.94 for the $D_{2,2}$ disk, the accuracy of VLR is still relatively better than that of ILR. For situation that requires better refinement, the VLR is better suited than the ILR. 

The speedup of approximately two orders of magnitude by CUDA is significant in allowing the direct method to become a realistic option on GPUs. In astrophysical simulations, a single galaxy simulation may require $512\times 512$ grid zones with $10^4$ iterations. Under such conditions, the computational time required is about $10$ days ($10^4\times 8$ seconds). To reduce the computational time to $8$ hours, $12$ cards of $P100$ GPUs have to be deployed at once. Further improvement of the performance on GPU can be made through optimization of algorithms based on different GPU architectures as a followup to the calculations reported in this paper. 

For the case of uniform grid discretization,
the double sum of the potential function has been represented as a convolution form of the surface density and the fundamental kernel (Yen et al 2012). 
A direct calculation requires the computational complexity of $O(N^4)$ if the number $N$ is the number of zones in one direction. The fast computation of Yen et al. (2012) can reduce the complexity $O(N^4)$ to nearly linear $O((N \log_2 N)^2)$ with the help of FFT. Such a comparison with the studies of multi-grid and spectral method has been given in Yen et al. (2012) for the uniform grid.
Here in this work, we compare the current approach that has been built on the extension of grid structure from the uniform discretization to one with adaptive mesh refinement. 
This generalization has been shown to preserve the order of accuracy successfully. The FFT technique can not be employed here to reduce computational time for AMR.

In this research, it should be emphasized that the formula $K^{x,0}_i$ should be modified for AMR. To reduce the computational time (Belleman et al 2008), 
various directions can still be explored. The most direct is to improve the availability of powerful computational facility. Building a tree structure and implementing an approach which is similar to the traditional $N$-body calculation will be considered next.
Finally, the framework of Cartesian coordinate calculation 
can also be generalized to Polar coordinates calculation,
following Yen et al. (2012) if the second order of accuracy can be relaxed.
The sub-rectangle 
$D_i=[x^{\ell}_i,x^r_i]\times [y^d_i,y^u_i]$ in Cartesian coordinates will be replaced by
the sub-rectangle $D_i=[r^{I}_i,r^O_i]\times [\theta^d_i,\theta^u_i]$
in Polar coordinates.
Since $\tilde r=1$ and $\theta=0$
can not hold simultaneously, 
the function $F(\tilde r,\theta)=\sqrt{1+{\tilde r}^2-2\tilde r\cos(\theta)}$
defined below Eqn (4.13) in Yen et al. (2012) is always positive for AMR. Hence the simple integrals 
\begin{eqnarray*}
\int \frac{\bar r(r-\bar r\cos(\theta))}{({\bar r}^2+r^2-2r{\bar r}\cos(\theta))^{3/2}}d{\bar r}
=-\cos(\theta)\ln \left(-\cos(\theta)+\frac{{\bar r}}{r}+F(\frac{{\bar r}}{r},\theta)\right)
+\frac{2\cos(\theta)\frac{\bar r}{r}-1}{F(\frac{\bar r}{r},\theta)}+C
\end{eqnarray*}
below Eqn (4.13) of Yen et al. (2012) and the term $-\cos(\theta)+\frac{{\bar r}}{r}+F(\frac{{\bar r}}{r},\theta)$ in the logarithmic function on the right hand side
$-\cos(\theta)+\frac{{\bar r}}{r}+F(\frac{{\bar r}}{r},\theta)$
should be modified as well for $\theta=0$ and $2|\frac{{\bar r}}{r}-1|$, to avoid singularity.
Other portions will be straightforward for derivation.

\begin{figure}[h]
\centerline{ \includegraphics[scale=.25]{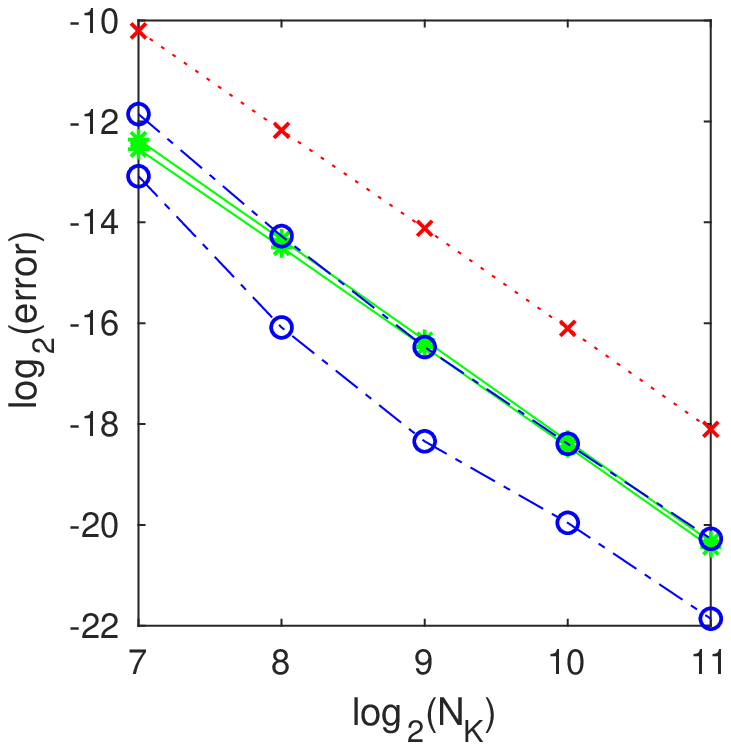}\quad \includegraphics[scale=.25]{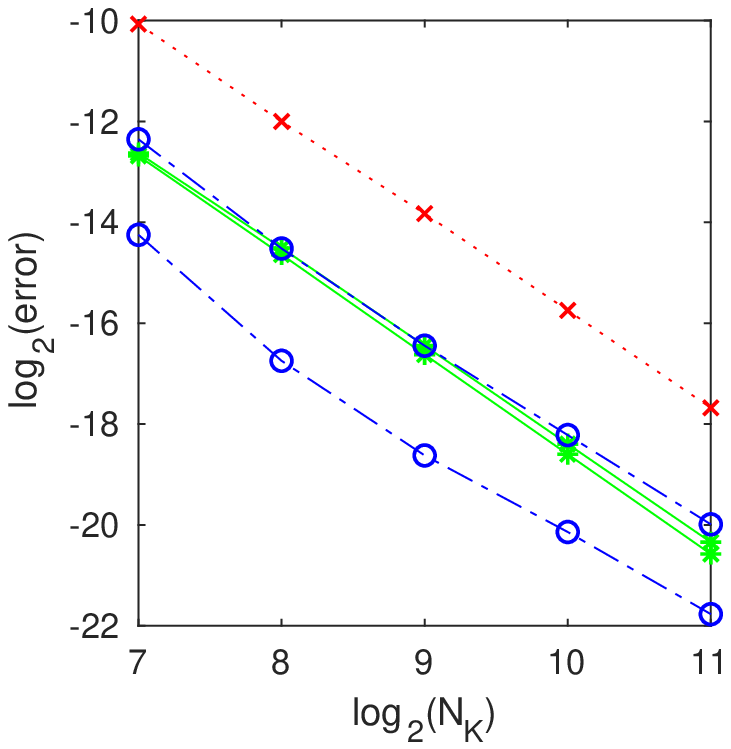}\quad \includegraphics[scale=.25]{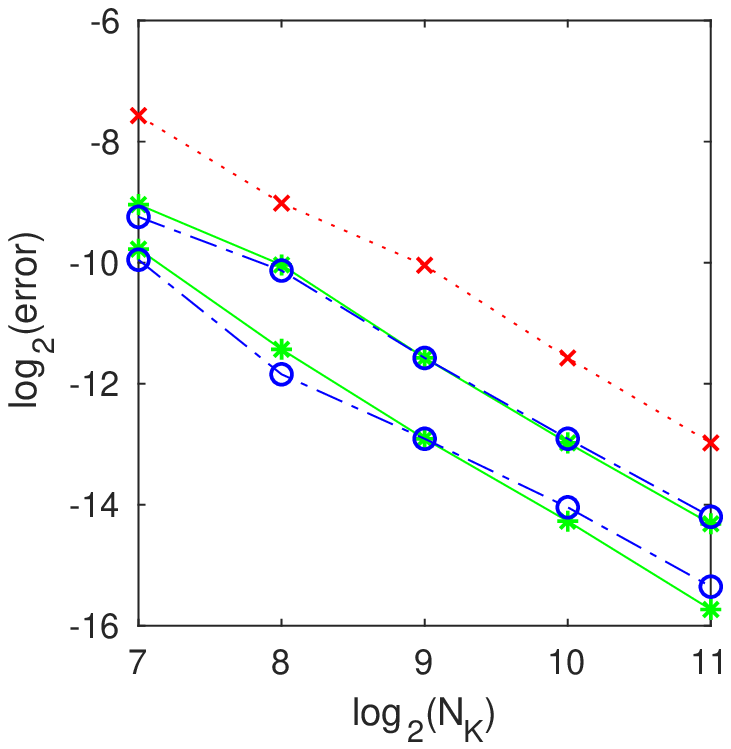} }
\centerline{ \includegraphics[scale=.25]{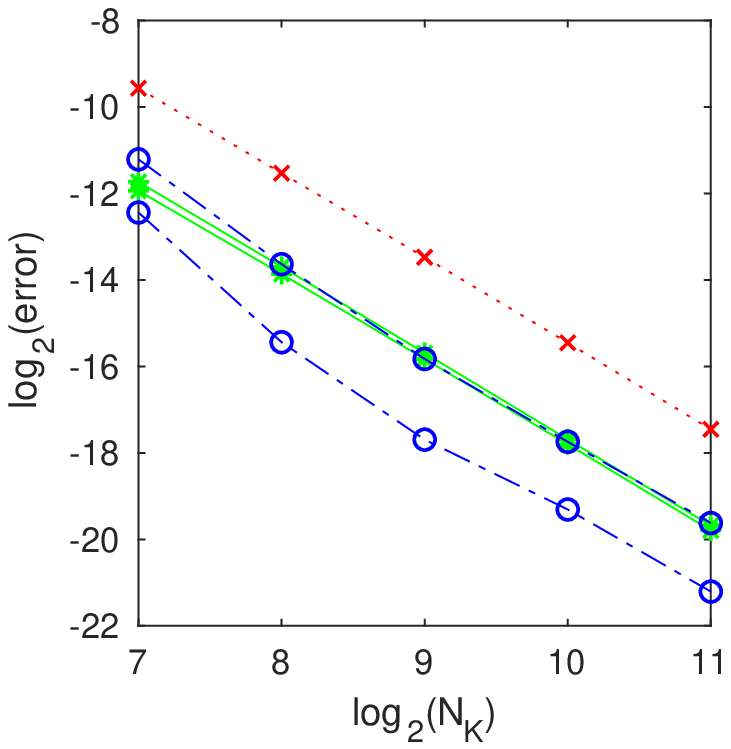}\quad \includegraphics[scale=.25]{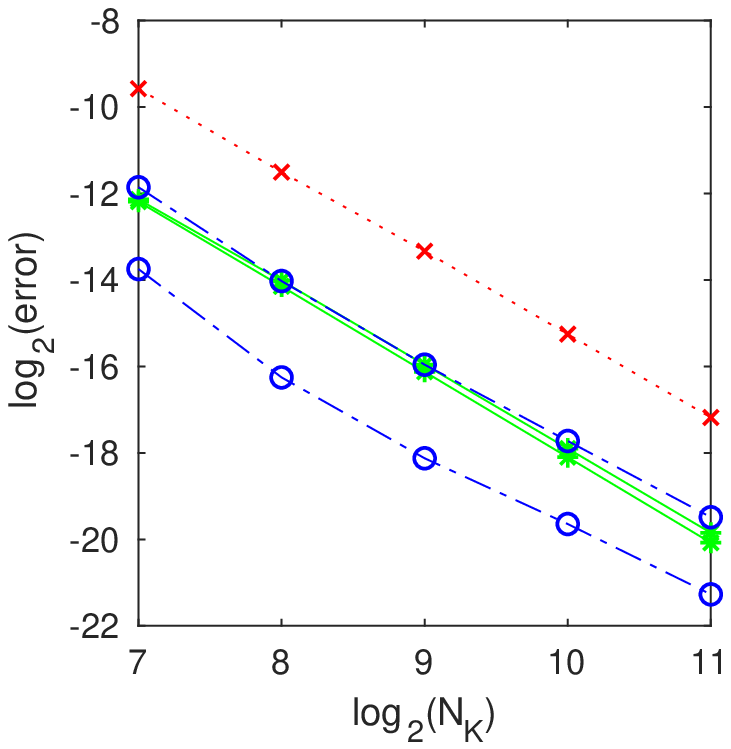}\quad \includegraphics[scale=.25]{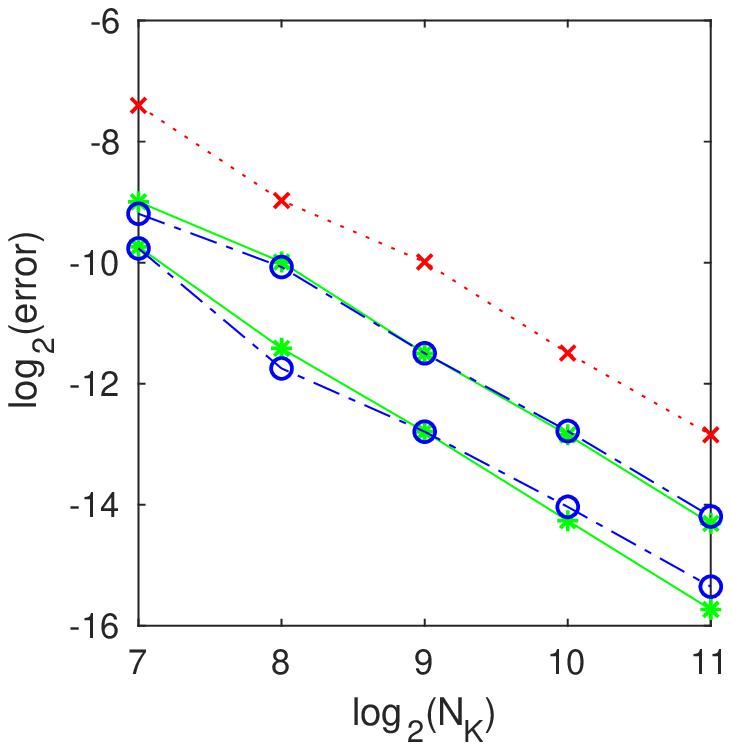} }
\caption{The behavior of accuracy for AMR($L$, $D_2$) simulations. The dotted, dashed, and solid lines represent the level $L=1,2,3$, respectively. TH
The symbols , ``x". ``*'' and ``o'' are for the cases of uniform grids, the AMR-ILR and AMR-VLR, respectively. The accuracy are shown in the $1$, $2$, and $\infty$-norms in the panels from left to right. The slopes of the fitted lines give the order of accuracy while their positions indicate their respective accuracy. The top row shows the force calculations in the $x$-direction, and the bottom row is for the radial force calculations. It shows that they have the same order of accuracy, but the accuracy from the $L=3$ refinement is the best.}
\end{figure}

\begin{figure}[h]
\centerline{ \includegraphics[scale=.25]{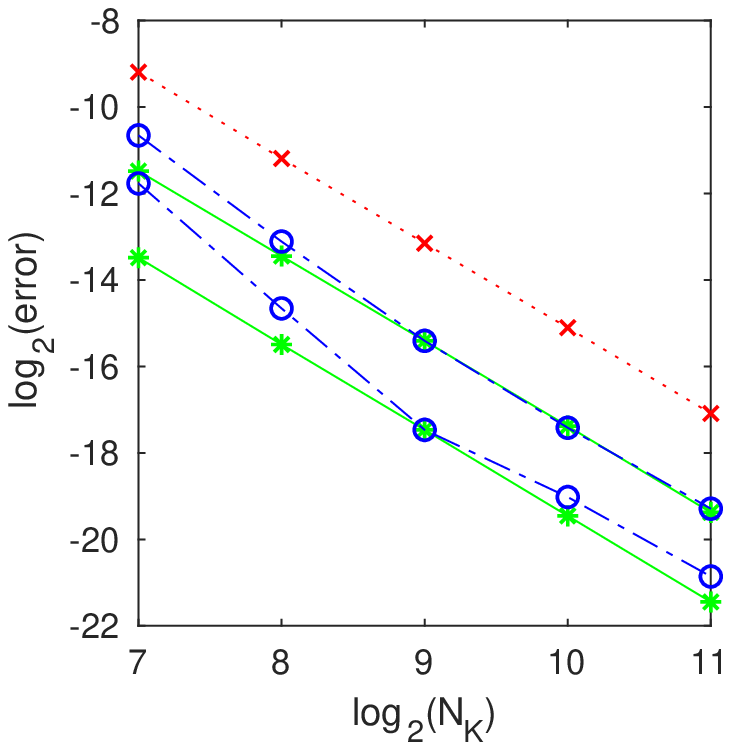}\quad \includegraphics[scale=.25]{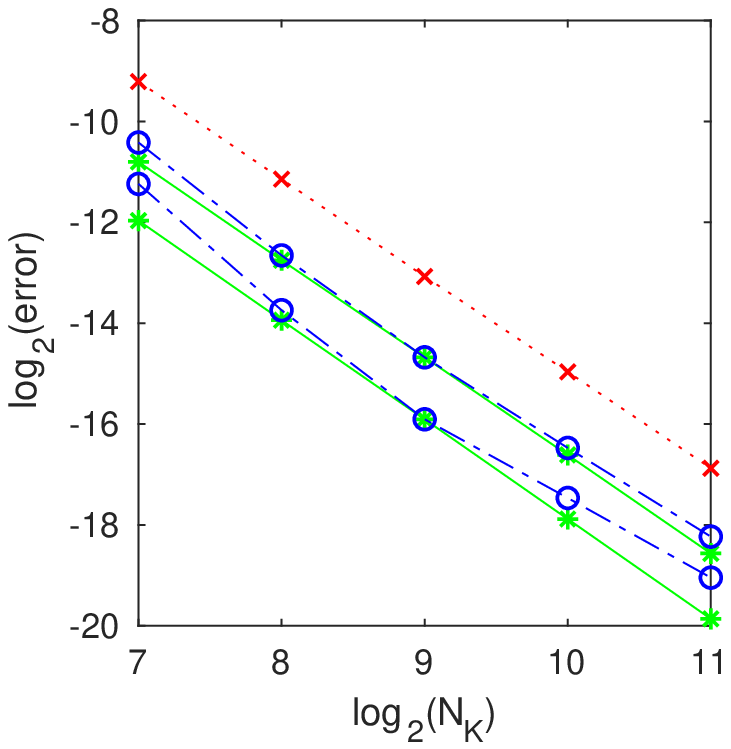}\quad \includegraphics[scale=.25]{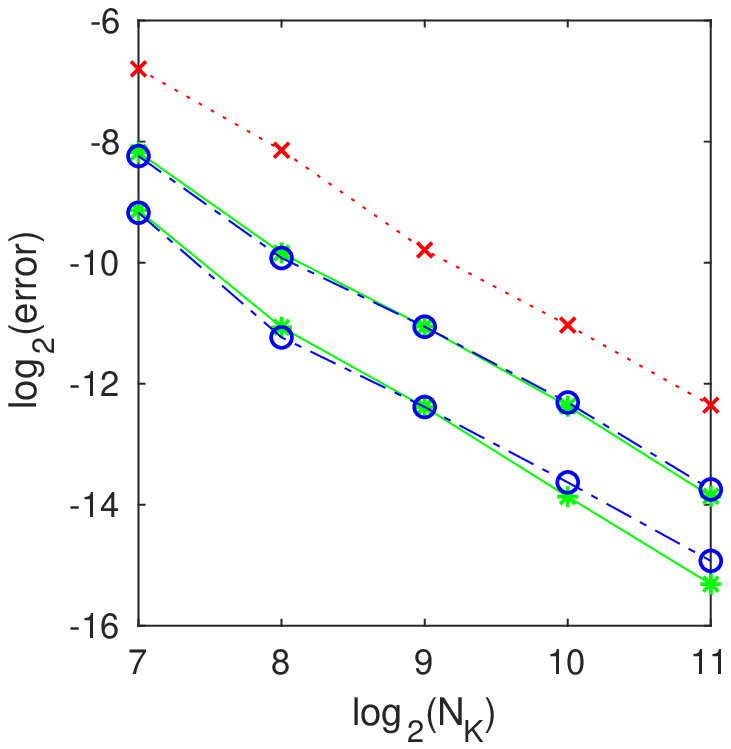} }
\centerline{ \includegraphics[scale=.25]{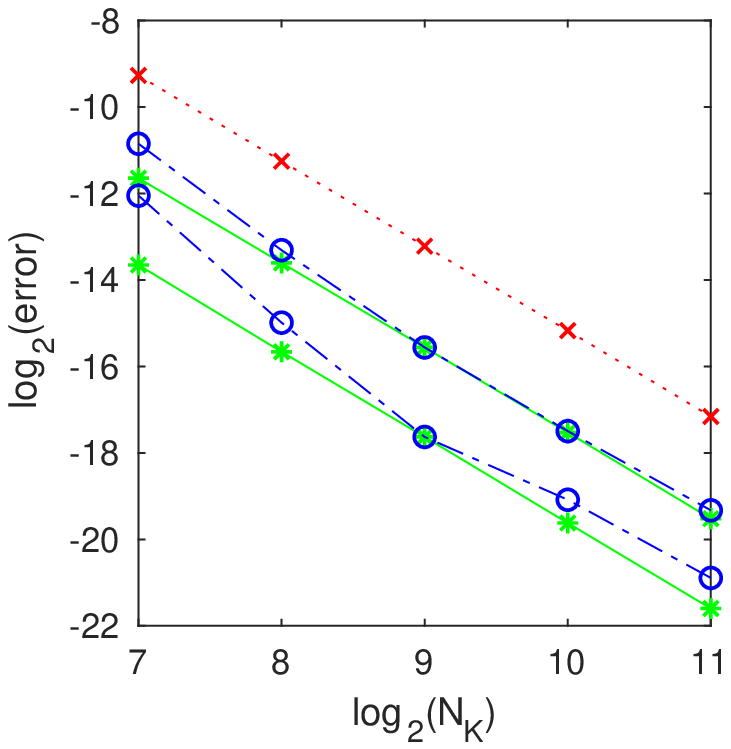}\quad \includegraphics[scale=.25]{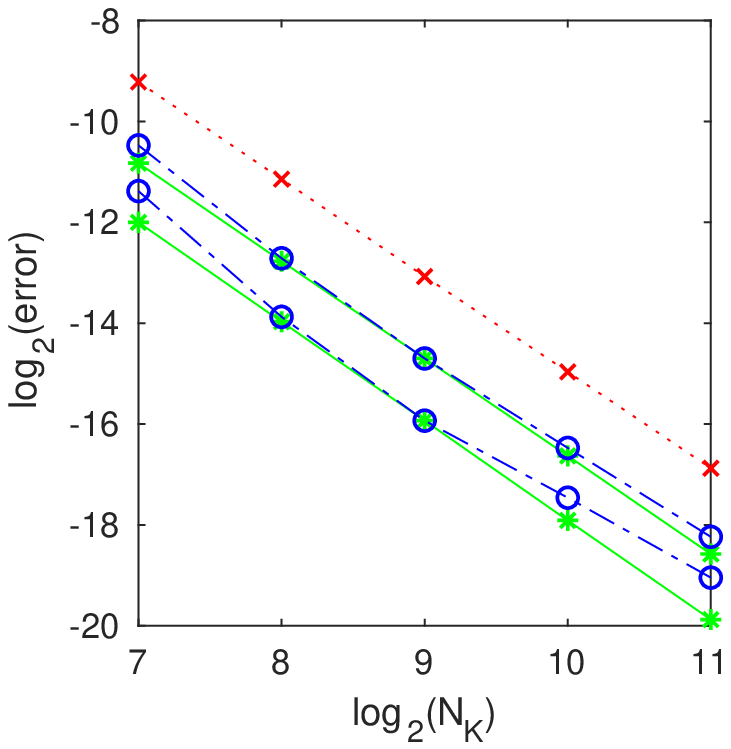}\quad \includegraphics[scale=.25]{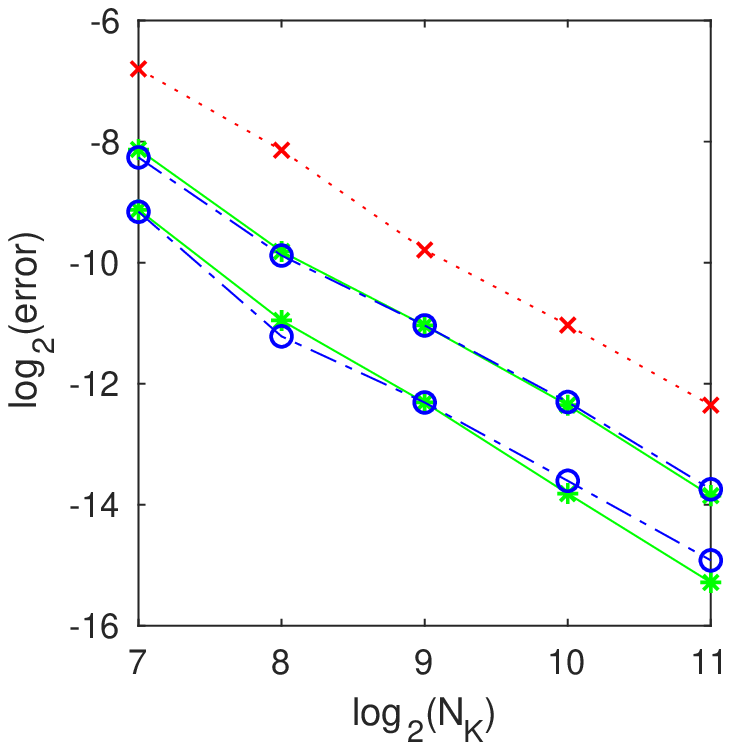} }
\centerline{ \includegraphics[scale=.25]{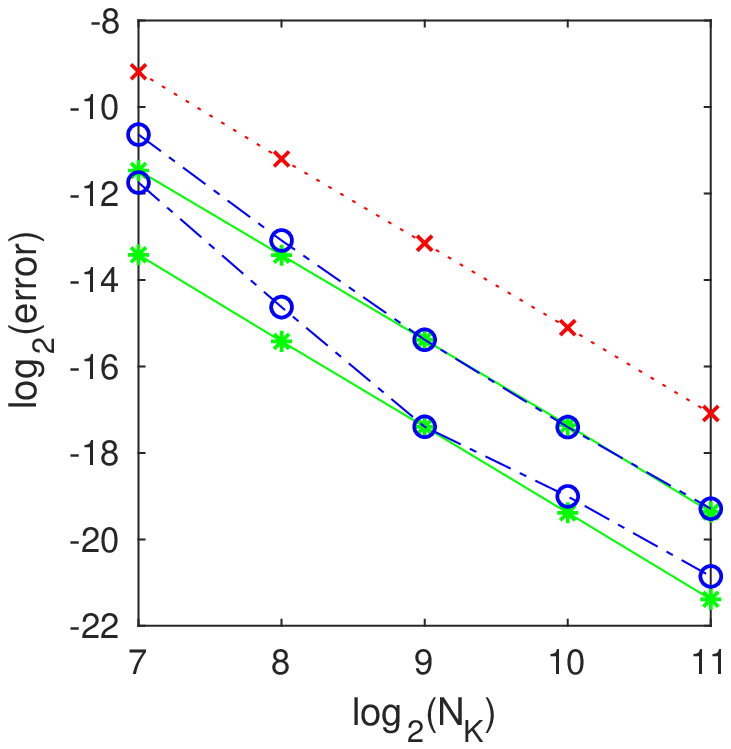}\quad \includegraphics[scale=.25]{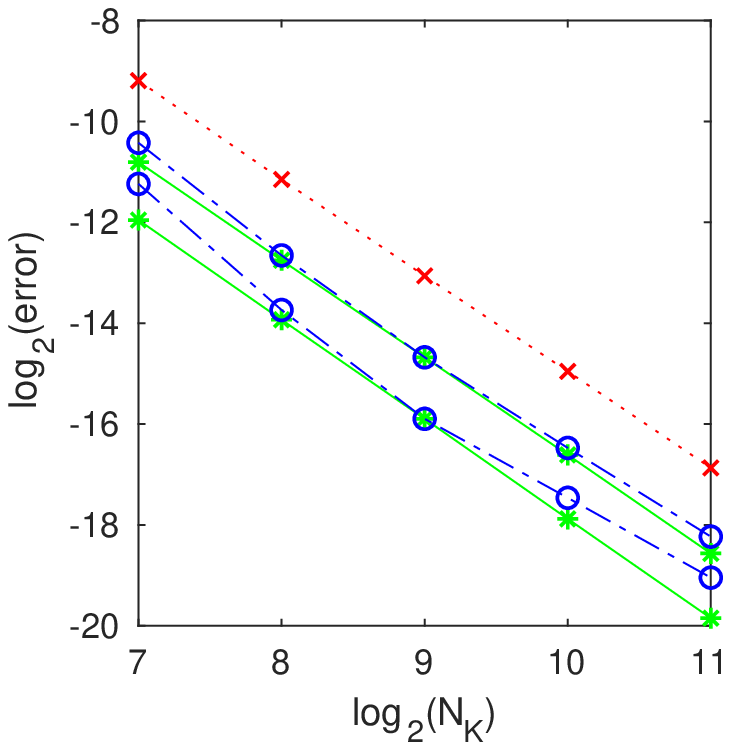}\quad \includegraphics[scale=.25]{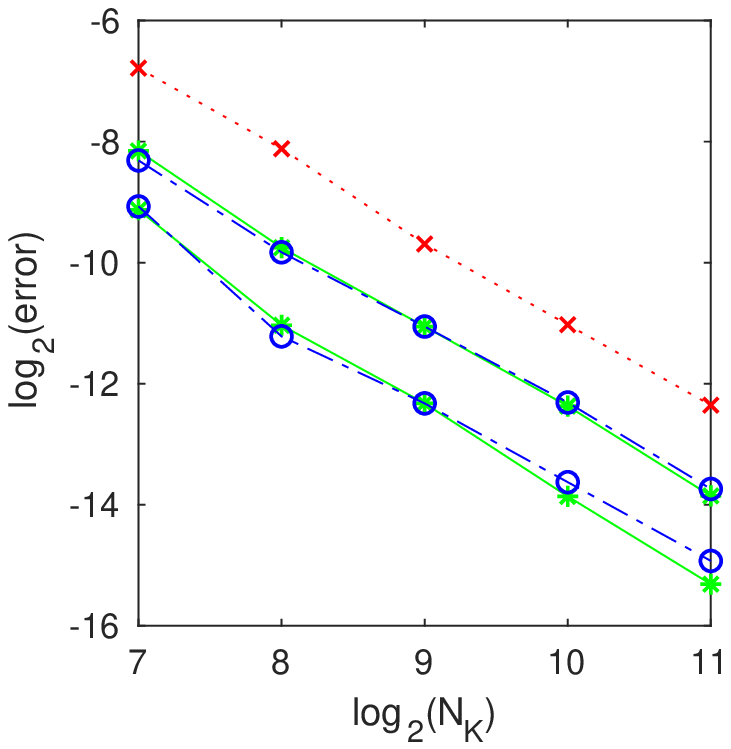} }
\caption{The behavior of accuracy for AMR($L$, $D_{2,2}$) simulations. 
The dotted (red), dashed (blue), and solid (cyan) lines label the level of refinement $L=1,2,3$, respectively. 
The symbols ``x", ``*'' and ``o'' label the uniform grid, AMR-ILR and AMR-VLR, respectively.
The accuracy of the $1$, $2$, and $\infty$-norms is shown in the columns from left to right. Slopes of the respective lines represent the order of accuracy. From top to bottom rows: $x$-direction force, $y$-direction force, and the radial force.
}
\end{figure}

\begin{figure}[ht]
\centerline{ 
\includegraphics[scale=.05]{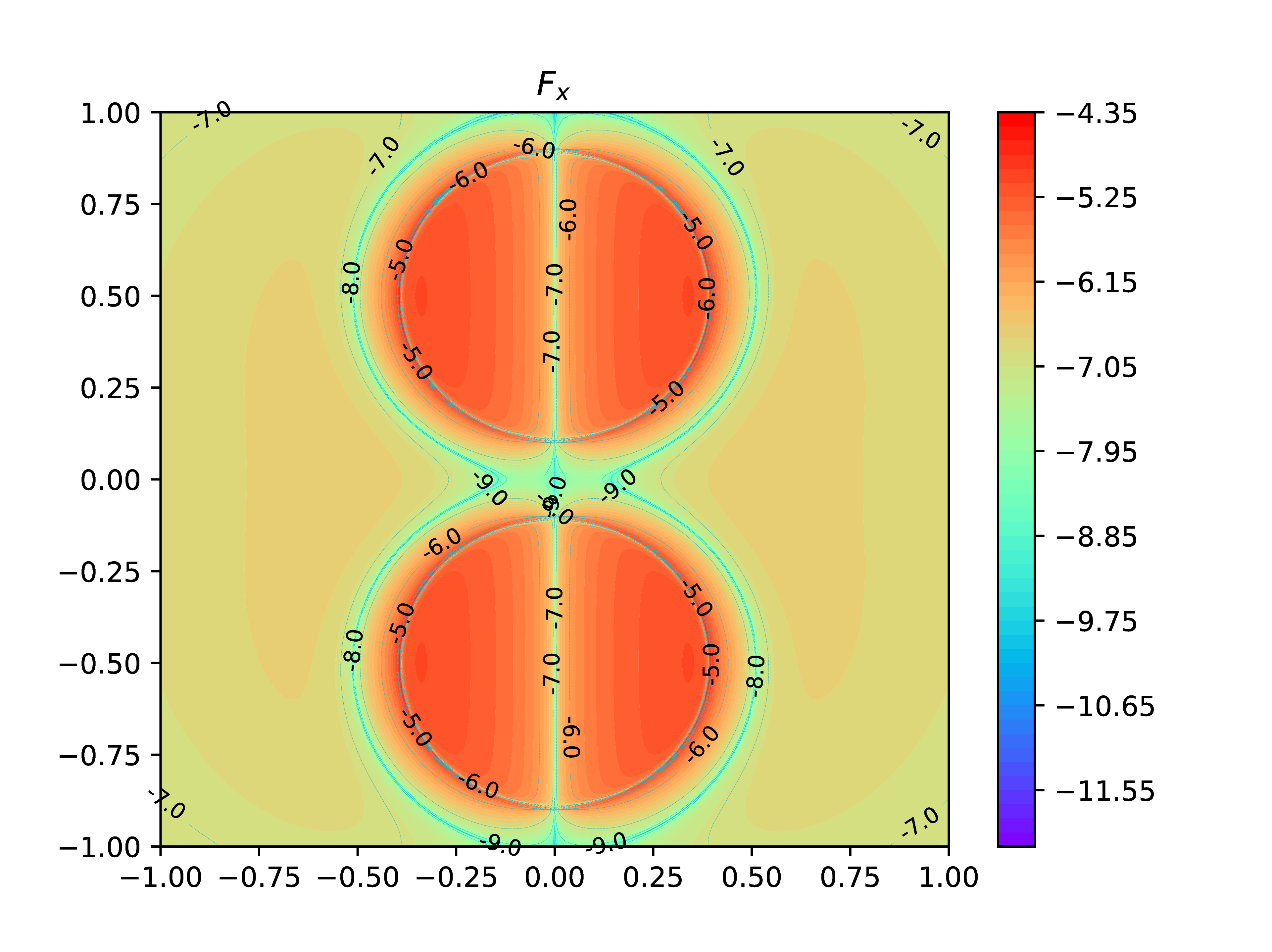}\thinspace 
\includegraphics[scale=.05]{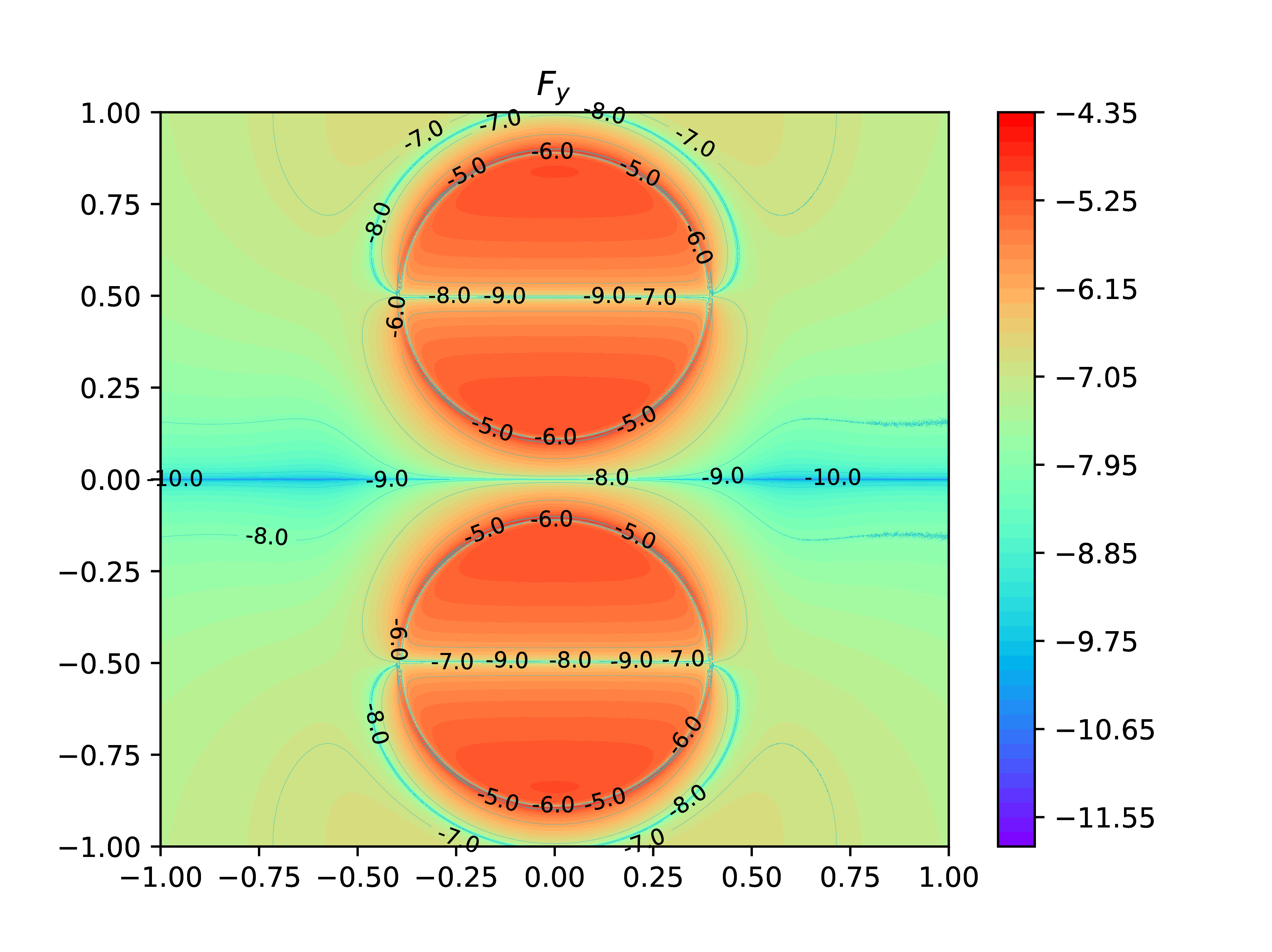}\thinspace 
\includegraphics[scale=.05]{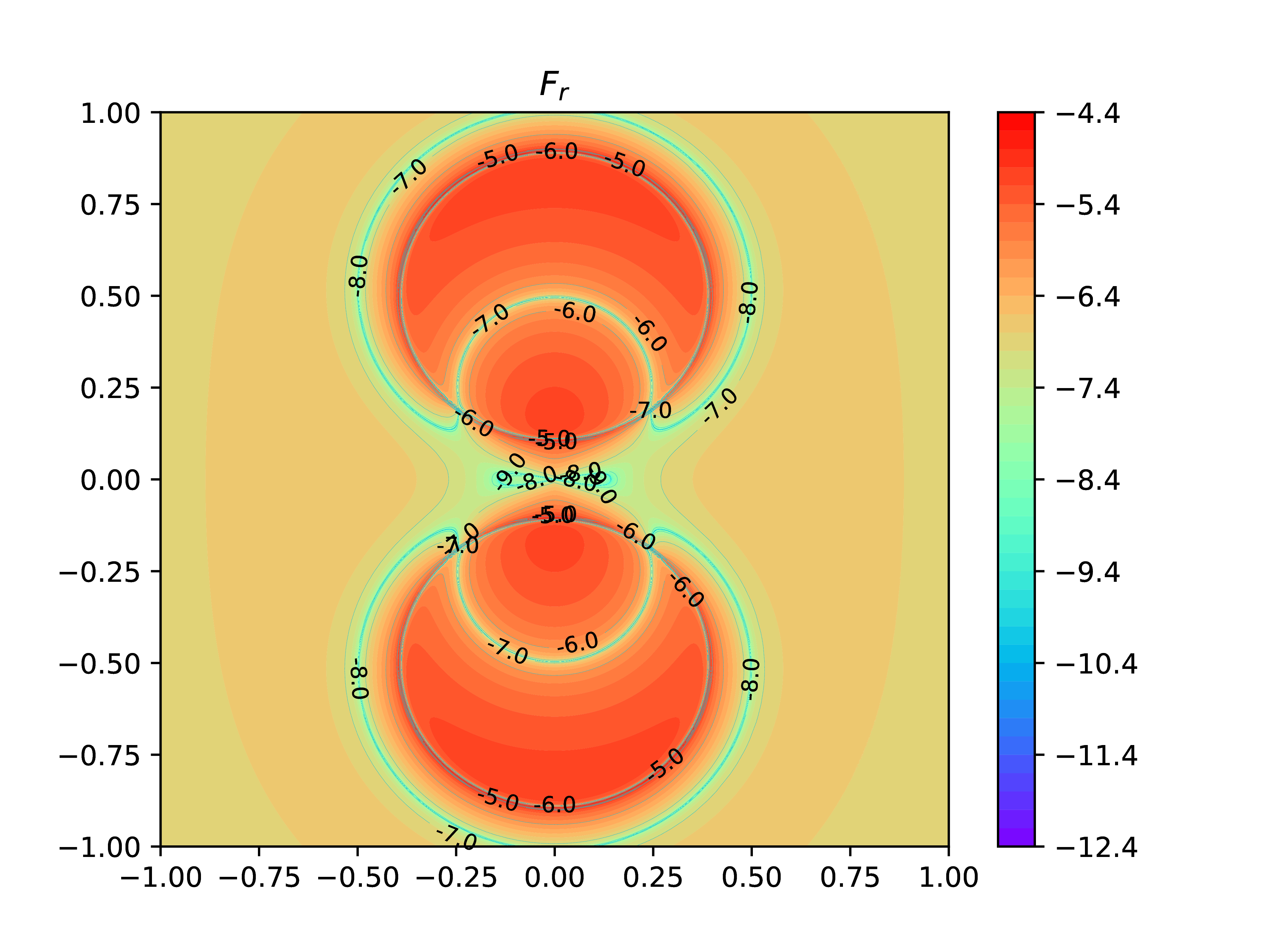} }
\caption{Contours of errors in $x$, $y$ and $R$ directions (from left to right) for the AMR(3, $D_{2,2}$) simulations are shown with ILR using $N_k=2048$. Larger errors occur near the edges of the disks.
}
\end{figure}

\acknowledgements
%\yhtseng{
The authors acknowledge the support of the Theoretical Institute for Advanced Research in Astrophysics (TIARA) based in Academia Sinica Institute of Astronomy and Astrophysics (ASIAA). C.C.Y. thanks the Institute of Astronomy and Astrophysics, Academia Sinica, Taiwan for their constant support. C.C.Y. is supported by Ministry of Science and Technology of Taiwan, under grant MOST-107-2115-M- 030-005-MY2. 
%}

\appendix

\section{Data of AMR according to Figure 1. }
\begin{table}[ht]
\begin{center}
\begin{tabular}{|c|c|c|c|c|c|c|c|c|c|}
\hline
ID & $x_c$ & $y_c$ & L & P & EL & LN & DN & RN & UN \\ \hline
1 & 0.13 & 0.13 & 1 & 0 & 1 & -1 & -1 & 2 & 5\\ \hline
2 & 0.38 & 0.13 & 1 & 0 & 1 & 1 & -1 & 3 & 6\\ \hline
3 & 0.63 & 0.13 & 1 & 0 & 1 & 2 & -1 & 4 & 7\\ \hline
4 & 0.88 & 0.13 & 1 & 0 & 1 & 3 & -1 & -1 & 8\\ \hline
5 & 0.13 & 0.38 & 1 & 0 & 1 & -1 & 1 & 6 & 9\\ \hline
6 & 0.38 & 0.38 & 1 & 0 & 0 & 5 & 2 & 7 & 10\\ \hline
7 & 0.63 & 0.38 & 1 & 0 & 1 & 6 & 3 & 8 & 11\\ \hline
8 & 0.88 & 0.38 & 1 & 0 & 1 & 7 & 4 & -1 & 12\\ \hline
9 & 0.13 & 0.63 & 1 & 0 & 1 & -1 & 5 & 10 & 13\\ \hline
10 & 0.38 & 0.63 & 1 & 0 & 1 & 9 & 6 & 11 & 14\\ \hline
11 & 0.63 & 0.63 & 1 & 0 & 0 & 10 & 7 & 12 & 15\\ \hline
12 & 0.88 & 0.63 & 1 & 0 & 1 & 11 & 8 & -1 & 16\\ \hline
13 & 0.13 & 0.88 & 1 & 0 & 1 & -1 & 9 & 14 & -1\\ \hline
14 & 0.38 & 0.88 & 1 & 0 & 1 & 13 & 10 & 15 & -1\\ \hline
15 & 0.63 & 0.88 & 1 & 0 & 1 & 14 & 11 & 16 & -1\\ \hline
16 & 0.88 & 0.88 & 1 & 0 & 1 & 15 & 12 & -1 & -1\\ \hline
17 & 0.31 & 0.44 & 2 & 6 & 0 & 5 & 19 & 18 & 10\\ \hline
18 & 0.31 & 0.31 & 2 & 6 & 1 & 17 & 20 & 7 & 10\\ \hline
19 & 0.44 & 0.31 & 2 & 6 & 0 & 5 & 2 & 20 & 17\\ \hline
20 & 0.44 & 0.44 & 2 & 6 & 1 & 19 & 2 & 7 & 18\\ \hline
21 & 0.56 & 0.69 & 2 & 11 & 1 & 10 & 23 & 22 & 15\\ \hline
22 & 0.56 & 0.56 & 2 & 11 & 1 & 21 & 24 & 12 & 15\\ \hline
23 & 0.69 & 0.56 & 2 & 11 & 1 & 10 & 7 & 24 & 21\\ \hline
24 & 0.69 & 0.69 & 2 & 11 & 1 & 23 & 7 & 12 & 22\\ \hline
25 & 0.28 & 0.47 & 3 & 17 & 1 & 5 & 27 & 26 & 10\\ \hline
26 & 0.28 & 0.41 & 3 & 17 & 1 & 25 & 28 & 18 & 10\\ \hline
27 & 0.34 & 0.41 & 3 & 17 & 1 & 5 & 19 & 28 & 25\\ \hline
28 & 0.34 & 0.47 & 3 & 17 & 1 & 27 & 19 & 18 & 26\\ \hline
29 & 0.41 & 0.34 & 3 & 19 & 1 & 5 & 31 & 30 & 17\\ \hline
30 & 0.41 & 0.28 & 3 & 19 & 1 & 29 & 32 & 20 & 17\\ \hline
31 & 0.47 & 0.28 & 3 & 19 & 1 & 5 & 2 & 32 & 29\\ \hline
32 & 0.47 & 0.34 & 3 & 19 & 1 & 31 & 2 & 20 & 30\\ \hline
\end{tabular}
\caption{Data of AMR according to Figure 1.} 
\end{center}
\end{table}

\end{document}